\documentclass[aps,pre,twocolumn,superscriptaddress,floatfix]{revtex4}

\usepackage{graphicx,bm,amssymb,amsmath,dcolumn,hyperref}
\usepackage{multirow}
\usepackage{enumerate}
\usepackage{enumitem}
\usepackage{color}
\usepackage{subfigure}  
\usepackage{epstopdf}
\usepackage{bbm}
\usepackage{graphicx}
\usepackage{hyperref}
\usepackage{threeparttable}
\usepackage{amsthm}
\usepackage{mathtools}
\usepackage{color}
\usepackage{tikz}
\usepackage{float}

\begin{document}
\newcommand{\be}{\begin{equation}}
\newcommand{\ee}{\end{equation}}
\newcommand{\half}{\frac{1}{2}}
\newcommand{\ith}{^{(i)}}
\newcommand{\im}{^{(i-1)}}
\newcommand{\gae}
{\,\hbox{\lower0.5ex\hbox{$\sim$}\llap{\raise0.5ex\hbox{$>$}}}\,}
\newcommand{\lae}
{\,\hbox{\lower0.5ex\hbox{$\sim$}\llap{\raise0.5ex\hbox{$<$}}}\,}

\definecolor{blue}{rgb}{0,0,1}
\definecolor{red}{rgb}{1,0,0}
\definecolor{green}{rgb}{0,1,0}
\newcommand{\blue}[1]{\textcolor{blue}{#1}}
\newcommand{\red}[1]{\textcolor{red}{#1}}
\newcommand{\green}[1]{\textcolor{green}{#1}}

\newcommand{\scrA}{{\mathcal A}}
\newcommand{\scrB}{{\mathcal B}}
\newcommand{\scrE}{{\mathcal E}} 
\newcommand{\scrF}{{\mathcal F}} 
\newcommand{\scrL}{{\mathcal L}}
\newcommand{\scrM}{{\mathcal M}} 
\newcommand{\scrN}{{\mathcal N}}
\newcommand{\scrS}{{\mathcal S}}
\newcommand{\scrs}{{\mathcal s}}
\newcommand{\scrP}{{\mathcal P}}
\newcommand{\scrO}{{\mathcal O}}
\newcommand{\scrR}{{\mathcal R}}
\newcommand{\scrC}{{\mathcal C}}
\newcommand{\scrV}{{\mathcal V}}
\newcommand{\scrD}{{\mathcal D}}
\newcommand{\dm}{d_{\rm min}}
\newcommand{\dx}{{\rm d}x}
\newcommand{\ds}{{\rm d}s}
\newcommand{\dt}{{\rm d}t}
\newcommand{\yh}{y_{\rm h}}
\newcommand{\yt}{y_{\rm t}}
\newcommand{\dc}{d_{\rm c}}
\newcommand{\pc}{p_{\rm c}}
\newcommand{\rhojunction}{\rho_{\rm j}}
\newcommand{\rhojunctionLim}{\rho_{{\rm j},0}}
\newcommand{\rhobranch}{\rho_{\rm b}}
\newcommand{\rhobranchLim}{\rho_{{\rm b},0}}
\newcommand{\rhononbridge}{\rho_{\rm n}}
\newcommand{\rhononbridgeLim}{\rho_{{\rm n},0}}
\newcommand{\percolationCluster}{C}
\newcommand{\leafFreeCluster}{C_{\rm \ell f}}
\newcommand{\bridgeFreeCluster}{C_{\rm bf}}

\newcommand{\PercolationSector}{\delta_{\rm P}}
\newcommand{\IsingSector}{\delta_{\rm I}}

\newcommand{\PP}{\mathbb{P}}
\newcommand{\EE}{\mathbb{E}}



\title{Percolation effects in the Fortuin-Kasteleyn Ising model on the complete graph}
\date{\today}
\date{\today}
\author{Sheng Fang}
\affiliation{Hefei National Laboratory for Physical Sciences at Microscale and 
	Department of Modern Physics, University of Science and Technology of China, 
	Hefei, Anhui 230026, China}
\author{Zongzheng Zhou}
\email{eric.zhou@monash.edu}
\affiliation{ARC Centre of Excellence for Mathematical and Statistical Frontiers (ACEMS),
	School of Mathematics, Monash University, Clayton, Victoria 3800, Australia}
\author{Youjin Deng}
\email{yjdeng@ustc.edu.cn}
\affiliation{Hefei National Laboratory for Physical Sciences at Microscale and 
	Department of Modern Physics, University of Science and Technology of China, 
	Hefei, Anhui 230026, China}
\affiliation{MinJiang Collaborative Center for Theoretical Physics,
College of Physics and Electronic Information Engineering, Minjiang University, Fuzhou 350108, China}

\begin{abstract}
The Fortuin-Kasteleyn (FK) random cluster model, which  can be exactly mapped from the $q$-state Potts spin model, is a correlated bond percolation model.
By extensive Monte Carlo simulations, we study the FK bond representation of the critical Ising model ($q=2$) on a finite complete graph, i.e. the mean-field Ising model.
We provide strong numerical evidence  that the configuration space for $q=2$ contains an asymptotically vanishing sector 
in which quantities exhibit the same finite-size scaling as in the critical uncorrelated bond percolation ($q=1$) on the complete graph.
Moreover,  we observe that in the full configuration space, the power-law behaviour of the cluster-size distribution 
for the FK Ising clusters except the largest one is governed by a Fisher exponent taking the value for $q=1$ instead of $q=2$.  
This demonstrates the percolation effects in the FK Ising model on the complete graph.
\end{abstract}
\maketitle

\section{Introduction}
\label{Introduction}
The Fortuin-Kasteleyn (FK) random cluster model~\cite{Grimmett2006} is a correlated bond percolation problem, defined by assigning the probability measure $\pi(A) \propto q^{k(A)}v^{|A|}$ to each bond configuration $A$. Here $k(A)$ is the number of connected components, or \emph{clusters}, on $A$. The parameter $v>0$ is the bond fugacity, and the cluster fugacity $q$ controls the preference to the number of clusters. Namely, the system prefers more (less) clusters if $q>1$ ($q<1$). When $q=1$, bonds become mutually independent and thus it simply corresponds to the standard (uncorrelated) bond percolation model, where the bond occupation probability $p$ relates to $v$ as $v=p/(1-p)$. 
The $q=2$ case is the FK Ising model, which can be mapped to the zero-field ferromagnetic Ising model via the FK transformation~\cite{kasteleyn1969phase}.

In statistical mechanics, models on the \emph{complete graph} (CG)~\footnote{A complete graph ${\rm K}_V$ is a graph with $V$ vertices, on which each vertex is adjacent to all others. are considered to describe the systems in the infinite-spatial-dimension limit and thus frequently referred to mean-field models.} The study of these models on the CG are of special interest, since it can provide a qualitative approximation and thus an insightful picture for finite dimensions.  
 In the context of phase transitions and critical phenomena, it is further expected that many quantities of these models on CG exhibit the same asymptotics as on the high-dimensional tori.
The random-cluster model on CG with general $q>0$ was first systematically studied in Ref.~\cite{BollobasGrimmettJanson1996}, and then extended to a broad family of pseudo-critical points in Ref.~\cite{LuczakLuczak2008}. 
In particular, the authors in Ref.~\cite{LuczakLuczak2008} proved that for the $q=2$ case,  the size of the largest cluster scales as $C_1 \sim V^{3/4}$ within the Ising critical window given by $1-p/p_{\rm c} = {\rm O}(V^{-1/2})$, where $p_{\rm c} = 2/V$ is the critical point and $V$ is the number of vertices on CG. The CG asymptotics for $C_1$ was later observed numerically on five-dimensional tori~\cite{lundow2015complete}.

In this paper, we study the FK Ising model on CG at the critical point, using extensive Monte Carlo simulations. We partition the bond configuration space into two sectors: $S_{\rm P}$ and $S_{\rm I}$, according to the size of the largest cluster. A bond configuration belongs to $S_{\rm P}$ if its size of the largest cluster is less than or equal to $2V^{2/3}$; otherwise it belongs to $S_{\rm I}$. The first message delivered by our data is that $\PP(S_{\rm P}) \sim V^{-\theta}$ with $\theta$ consistent $1/12$, and moreover, conditioned on $S_{\rm P}$, many quantities are observed to exhibit the same scaling as their analogous quantities in the critical uncorrelated percolation on CG. For instance, the size of the second largest cluster scales as $C_2^{\rm P}\sim V^{2/3}$, and the cluster-size distribution obeys the same scaling function as that for the percolation model on CG.  We thus call $S_{\rm P}$ a \emph{percolation sector}. The other sector $S_{\rm I}$ is called the Ising sector, conditioned on which we observe that $C^{\rm I}_1 \sim V^{3/4}$, and $C^{\rm I}_2\sim \sqrt{V}\log V$.

As a consequence, for an arbitrary observable $\mathcal{Q}$, its average can be written as
\begin{equation}
\label{Eq:Conjectured quantities scaling}
    \EE (\mathcal{Q}) = \EE(\mathcal{Q} | S_{\rm P})\PP(S_{\rm P}) + \EE(\mathcal{Q} | S_{\rm I})\PP(S_{\rm I}) \;.
\end{equation}
As mentioned above, our data suggest that $\PP(S_{\rm P}) \sim V^{-1/12}$ and $\PP(S_{\rm I}) \sim 1 - cV^{-1/12}$, with some positive constant $c$. For the largest cluster $\scrC_1$, we observed $C^{\rm P}_1 := \EE(\mathcal{C}_1 | S_{\rm P}) \sim V^{2/3}$ and $C^{\rm I}_1 := \EE(\mathcal{C}_1 | S_{\rm I}) \sim V^{3/4}$. Then using Eq.~\eqref{Eq:Conjectured quantities scaling} gives that $C_1 \sim V^{3/4} - a_1V^{2/3} + a_2V^{7/12}$ with some positive constants $a_1, a_2$. Similarly, the size of the second largest cluster $C_2 \sim V^{7/12} + b_1\sqrt{V}\log V - b_2V^{5/12}\log V$, with some positive constants $b_1, b_2$. As one can see, although the effect of the percolation sector to $C_1$ is subdominant, it dominates the scaling of $C_2$.

Moreover, we also observe percolation effects in the geometric properties of the FK Ising clusters, except the largest one, without conditioning on any special sector in the configuration space. Consider the cluster-size distribution $n(s,V)$. The standard finite-size scaling, as observed for the percolation model~\cite{huang2018critical} and the low-dimensional FK Ising model~\cite{hou2019geometric}, predicts that $n(s,V) \sim s^{-\tau}\tilde{n}(s/V^{d_{\rm f}})$, where $d_{\rm f}$ is the volume fractal dimension of the largest cluster ($C_1 \sim V^{d_{\rm f}}$) and $\tilde{n}(\cdot)$ is a universal scaling function. The Fisher exponent $\tau$ relates to $d_{\rm f}$ by the scaling relation as $\tau = 1 + 1/d_{\rm f}$. Substituting the fractal dimension $d_{\rm f} = 3/4$ for the CG Ising model gives $\tau = 7/3$, and substituting $d_{\rm f} = 2/3$ for the CG percolation gives $\tau = 5/2$. Surprisingly, our data clearly suggest $\tau=5/2$ instead of $7/3$, which demonstrate the percolation effects in the FK Ising model on the CG. Moreover, we also find that the scaling function $\tilde{n}(x) \approx 1/\sqrt{2\pi}$ when $x \ll 1$, again same as the observation in the percolation model on the complete graph~\cite{ben2005kinetic,huang2018critical}.
   
The remainder of this paper is organized as follows. Sec.~\ref{Algorithm and Simulation} summarizes the simulation details and observables. Sec.~\ref{Results} contains our main numerical results.  In Sec.~\ref{Sec:RG}, we provide an explanation to the percolation scaling window from the perspective of a renormalization group. A discussion is present in Sec.~\ref{Discussion}.

  \section{Simulations and Observables}  
  \label{Algorithm and Simulation}
  The algorithms we used to simulate the FK Ising model are Wolff and Swendsen-Wang (SW) algorithms~\cite{swendsen1987nonuniversal,wolff1989collective}. The update of SW algorithm involves two steps. For a given bond configuration, identify all connected components and for each of them, randomly and uniformly assign either $+$ or $-$ spins to all vertices on it. This step thus maps configurations from bonds to spins. Given a spin configuration, for each vertex, adding a bond between this vertex and its neighbours with probability $p = 1 - e^{-2K}$ if they have the same spin; otherwise do not add a bond. Given a spin configuration, the Wolff algorithm updates the spins as follows. Grow only one cluster from a uniformly and randomly chosen vertex, using the same connectivity rule as SW. Then generate a new spin configuration by flipping all spins on the cluster. 
  
  In two and three dimensions, it was numerically observed~\cite{wolff1989comparison} that the Wolff algorithm has smaller dynamic exponent than SW. Thus, in simulations, we used Wolff algorithm to update and decrease the correlations of spin configurations and only used the SW update to create clusters for sampling. The number of Wolff updates between two consecutive SW steps is chosen to be approximately the volume divided by the averaged size of clusters, such that every spin has a decent chance to be updated during these steps. In both Wolff and SW updates, to speed up the process of adding bonds between neighbouring vertices with the same spin, we adopt the procedure described in detail in~\cite{huang2018critical}.
  
In simulations, every time a bond configuration is generated by the SW step, we sample
\begin{enumerate}[label=(\alph*)]
    \item the size of every cluster, and we denote $\scrC_1, \scrC_2$ the sizes of the largest and second largest clusters;
    \item the cluster number $\scrN(s)$, defined as the number of clusters with size in $[s,s+\Delta s]$ with an appropriately chosen interval size $\Delta s$.
\end{enumerate}
We then define the percolation sector and the Ising sector as follows. A bond configuration $A$ is classified into the percolation sector $S_{\rm P}$ if and only if $\scrC_1(A)\leq 2V^{2/3}$; otherwise $A$ is classified into the Ising sector $S_{\rm I}$. Refer to Sec.~\ref{Anomalous scaling behavior of $C_1$} for the motivation of such a definition. We then study the ensemble average of the following quantities.
   	\begin{enumerate}[label=(\roman*)]
   	\item The averaged size of the largest and the second largest clusters: $C_1 = \langle \scrC_1 \rangle$, $C_2 = \langle	\scrC_2 \rangle$.
   	\item The cluster-size distribution $n(s,V) = \frac{1}{V\Delta s} \langle \scrN(s) \rangle$.
   	\item The size of the second largest cluster conditioned being in the percolation sector: $C^{\rm P}_2 = \langle \scrC_2|S_{\rm P}\rangle$.
   	\item The size of the second largest cluster conditioned being in the Ising sector: $C^{\rm I}_2 = \langle \scrC_2|S_{\rm I}\rangle$.
   	\item The cluster-size distribution density conditioned in the percolation sector: $n_{\rm P}(s,V) = \frac{1}{V\Delta s} \langle \scrN(s)|S_{\rm P} \rangle$.
   	\item The cluster-size distribution conditioned in the Ising sector: $n_{\rm I}(s,V) = \frac{1}{V\Delta s} \langle \scrN(s)|S_{\rm I} \rangle$.
   	\end{enumerate}
 In addition, using our recorded samples, we generate the probability density functions for $\scrC_1$, $\scrC_2$, $\scrC_2|S_{\rm P}$ and $\scrC_2|S_{\rm I}$.

  \section{Results}
  \label{Results} 
 
   \subsection{Anomalous scaling behavior of $\scrC_1$}
   \label{Anomalous scaling behavior of $C_1$} 
   
We first study $f_{\scrC_1}(s)$, the probability density function of the size of the largest cluster $\scrC_1$. Define $X_1 = \frac{\scrC_1}{V^{3/4}}$ and its probability density function as $f_{X_1}(x)$. Then it follows that
\begin{equation}
f_{\scrC_1}(s) \ds = f_{X_1}(x)\dx \;,
\end{equation}
where $\dx = V^{-3/4} \ds$, and thus $f_{X_1}(x) = V^{3/4}f_{\scrC_1}(s) $.

  \begin{figure}
    \centering
    \includegraphics[scale=0.8]{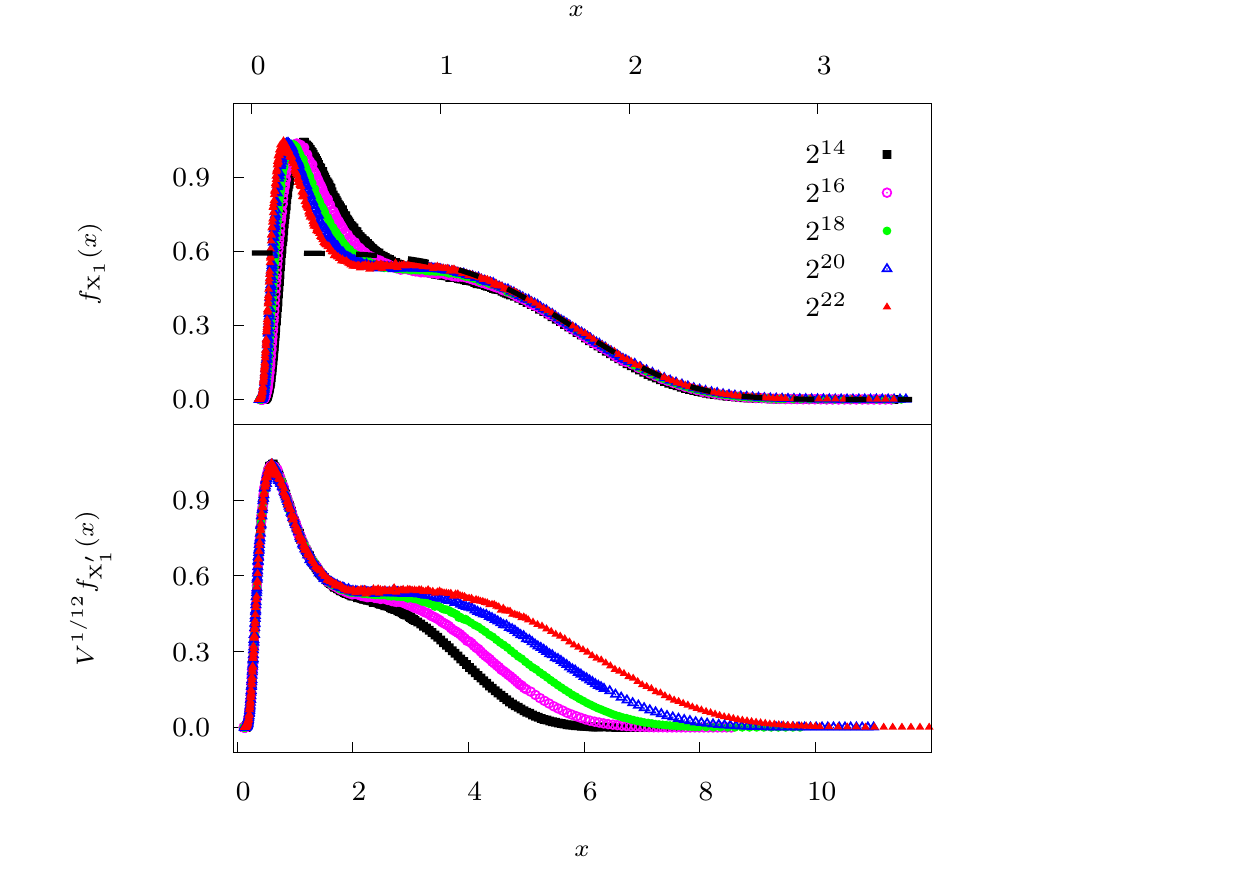}
    \caption{The probability distribution of $\scrC_1$, the size of the largest cluster. Here $f_{X_1}(x)$ (top) and $f_{X'_1}(x)$ are respectively the probability density functions of $X_1 = \scrC_1/V^{3/4}$ and $X'_1 = \scrC_1/V^{2/3}$. The black dashed line in the top figure shows the rigorous limiting distribution $f^{\infty}_{X_1}(x)$, shown in Eq.~\eqref{Eq:limiting distribution of C1}.}
      \label{Fig:distribution of C1}
  \end{figure}

It follows by Eq.(32) in~\citep{LuczakLuczak2008} that, as $V\rightarrow \infty$, $X_1$ has a limiting probability density function
\begin{equation}
\label{Eq:limiting distribution of C1}
f^{\infty}_{X_1}(x) = \frac{\exp(-x^4/12) }{\int_{0}^\infty \exp(-t^4/12) \dt }\;.
\end{equation}
In the top of Fig.~\ref{Fig:distribution of C1}, we plot $f_{X_1}(x)$ versus $x$ for various system sizes, and the data collapse to $f^{\infty}_{X_1}(x)$ when $x$ is approximately larger than $1$. For smaller $x$, each system shows a single-peak distribution, and this region fails to collapse for finite systems. Our data also imply that the area of this region tends to shrink and the locations of the peaks are approaching zero as $V$ goes to infinity. It is expected that in the $V\rightarrow \infty$ limit, our data of $f_{X_1}(x)$ converge to the limiting distribution $f^{\infty}_{X_1}(x)$.

The observation for the small $x$ region in the top of Fig.~\ref{Fig:distribution of C1} suggests that there might exist an asymptotically-decaying sector in which $\scrC_1$ is over-scaled by $V^{3/4}$, namely conditioned on being in this sector $C_1\sim V^y$ with some $y < 3/4$. It was proved in~\cite{LuczakLuczak2008} that, for the FK Ising model on the complete graph, there is a percolation scaling window near the critical point. So we conjecture that in this sector $C_1 \sim V^{2/3}$. Define $X_1' = \scrC_1/V^{2/3}$ and its corresponding probability density function $f_{X'_1}(x)$. A preliminary plot shows that $f_{X'_1}(x)$ multiplied by $V^{\theta}$ with $\theta \approx 1/12$ exhibits good data collapse for various system sizes when $x < 2$. This seemingly suggests that there exist an exponent $\theta$ and some positive constants $a_0,c_0$ such that
\begin{equation}
\lim_{V\rightarrow \infty} V^{\theta} \PP\left[\frac{\scrC_1}{V^{2/3}} \leq a_0\right]  = c_0 \;,
\end{equation}
$c_0 = \int_{0}^{a_0} f_{X'_1}(x) \dx$. To precisely estimate the exponent $\theta$, we calculate the probability of bond configurations whose $\scrC_1/V^{2/3} \in (1/2, 3/2)$. We perform least-squares fits on the data of this probability to the ansatz $c_0 V^{\theta}$.  
Our fits give that $\theta= -0.0833(4)$; we therefore conjecture the exact value of $\theta$ is $-1/12$. An interesting observation is that the conjectured value of $\theta$ is simply the difference between $2/3$ and $3/4$, the fractal dimensions of the largest cluster in the percolation model and the FK Ising model, respectively. The plot of $f_{X'_1}(x)V^{1/12}$ versus $x$ for various system sizes is shown in the bottom of Fig.~\ref{Fig:distribution of C1}.
        
\begin{figure}
    \centering
    \includegraphics[scale=0.6]{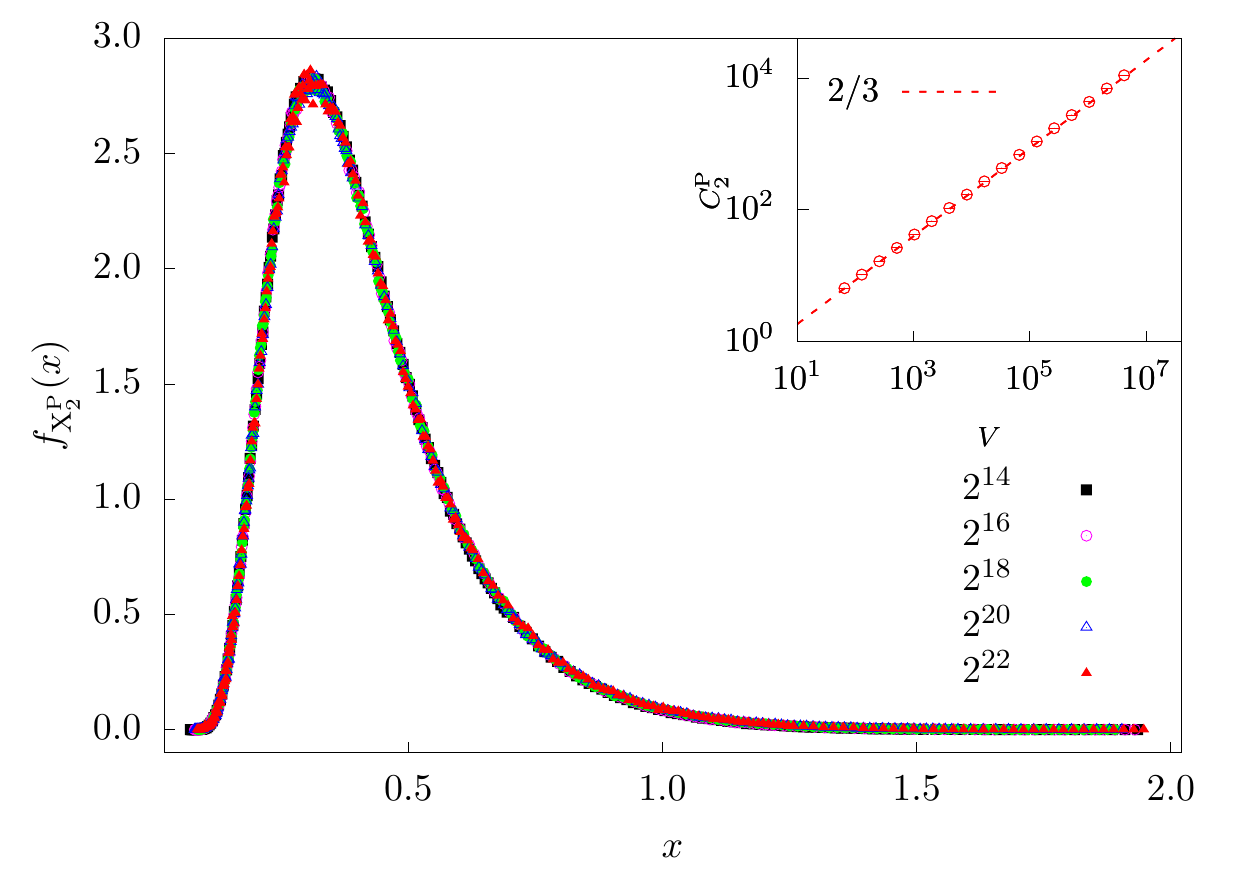}
    \caption{The probability distribution of $\scrC^{\rm P}_2$, the size of the second largest cluster in the percolation sector. Here $f_{X^{\rm P}_2}(x)$
is the probability density function of $X^{\rm P}_2 = \scrC^{\rm P}_2/V^{2/3}$. The inset plots the average $C^{\rm P}_2$ versus $V$.}
    \label{Fig:distribution of C2 in percolation sector}
  \end{figure}
  
\subsection{Further evidence for the percolation sector}
\label{Sec:Further evidence for the percolation sector}
In the above section, our data detect a sector which decays asymptotically with a rate $V^{-1/12}$. We numerically found that in this sector $C_1 \sim V^{2/3}$, which is consistent with the scaling of the largest cluster in the critical percolation model on the complete graph~\cite{nachmias2008critical}.

  \begin{figure}
    \centering
    \includegraphics[scale=0.6]{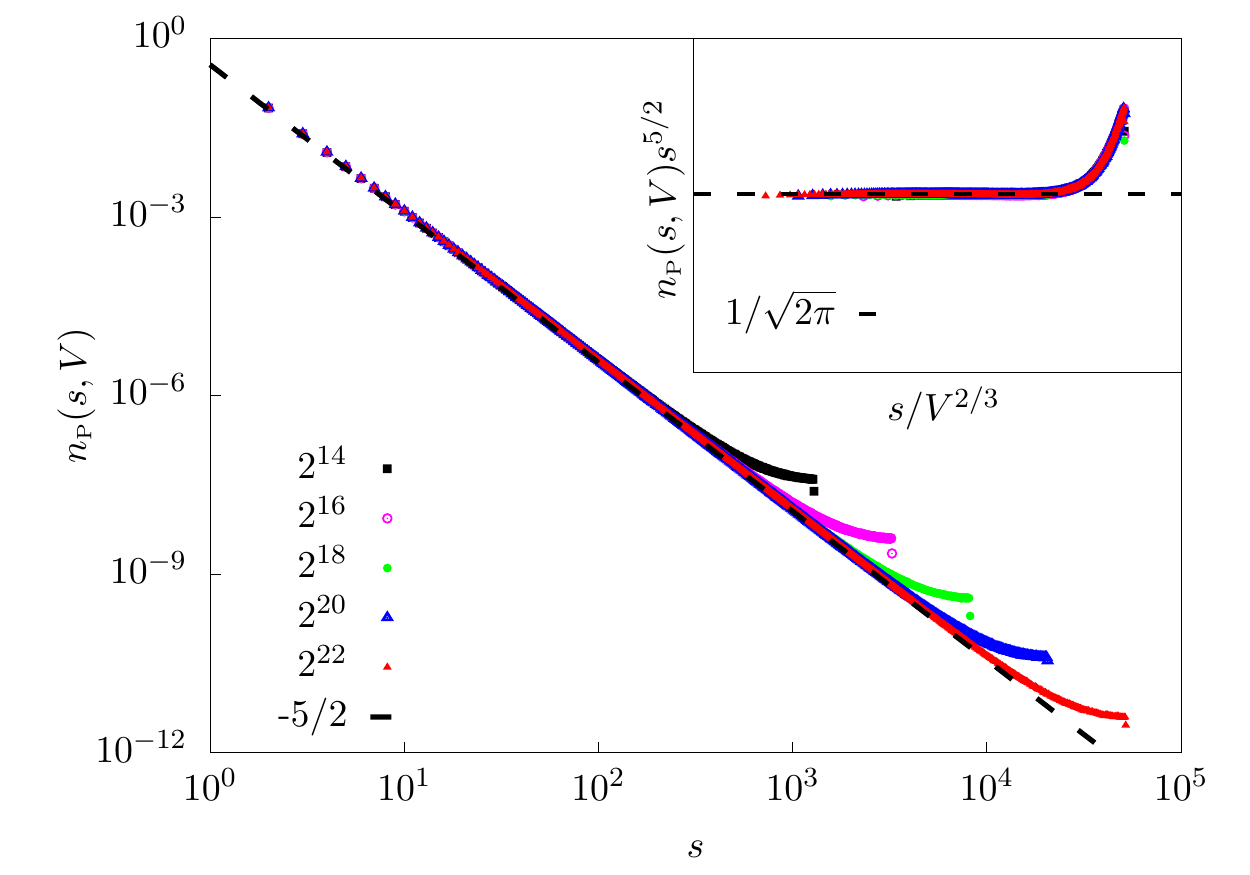}
    \caption{The cluster-size distribution $n_{\rm P}(s,V)$ in the percolation sector. The slope of the black dashed line is $-5/2$. The inset plots in log-log scale $n_{\rm P}(s, V) s^{5/2}$ versus $s/V^{2/3}$. It shows the scaling function $\tilde{n}(x)$ is consistent with $1/\sqrt{2\pi}$ if $x \ll 1$.}
    \label{Fig:cluster-size distribution in percolation sector}
  \end{figure}

To provide further evidence of that such a sector is percolation, we study $\scrC^{\rm P}_2 := \scrC_2|S_{\rm P}$, the size of the second largest cluster conditioned on that it is in the percolation sector. For critical percolation on the complete graph, it has been proved~\cite{LuczakLuczak2008} that the size of the second largest cluster also scales as $V^{2/3}$. We thus expect to observe the same scaling for $C^{\rm P}_2$. Define $X^{\rm P}_2:= \scrC^{\rm P}_2 / V^{2/3}$ and its probability density function $f_{X^{\rm P}_2}(x)$. In Fig.~\ref{Fig:distribution of C2 in percolation sector}, we plot $f_{X^{\rm P}_2}(x)$ for various system sizes and indeed good data collapse are observed. As expected, our data imply $C^{\rm P}_2 \sim V^{2/3}$, shown as the inset of Fig.~\ref{Fig:distribution of C2 in percolation sector}.

We then study $n_{\rm P}(s, V)$, the cluster-size distribution in the percolation sector. For critical percolation on the complete graph, it was numerically observed~\cite{huang2018critical} that $n(s, V) \sim s^{-\tau}\tilde{n}(s/V^{d_{\rm f}})$ where the $d_{\rm f} = 2/3$ is the fractal dimension of the largest cluster, and the Fisher exponent relates to $d_{\rm f}$ as
\begin{equation}
\label{Eq:Fisher exponent}
    \tau = 1 + 1/d_{\rm f}
\end{equation}
So, one has $\tau = 5/2$ for critical percolation on the complete graph. The scaling function $\tilde{n}(x)$ is approximately a constant if $x\ll 1$ and decays quickly to zero if $x\gg 1$. We conjecture $n_{\rm P}(s, V)$ exhibits the same scaling as $n(s, V)$. In Fig.~\ref{Fig:cluster-size distribution in percolation sector}, we plot the data for $n_{\rm P}(s, V)$ in log-log scale. As expected, it shows a clear power-law behaviour with the exponent consistent with $5/2$. Moreover, to show the scaling function, we log-log plot $n_{\rm P}(s, V) s^{5/2}$ versus $s/V^{2/3}$, shown as the inset of Fig.~\ref{Fig:cluster-size distribution in percolation sector}. As can be seen, the scaling function is consistent with the constant $1/\sqrt{2\pi}$ when $s/V^{2/3} \ll 1$. This is also the same constant observed in the critical percolation model on the complete graph~\cite{ben2005kinetic}.

\subsection{Ising sector}
\label{Sec:Ising}
  \begin{figure}
    \centering
    \includegraphics[scale=0.65]{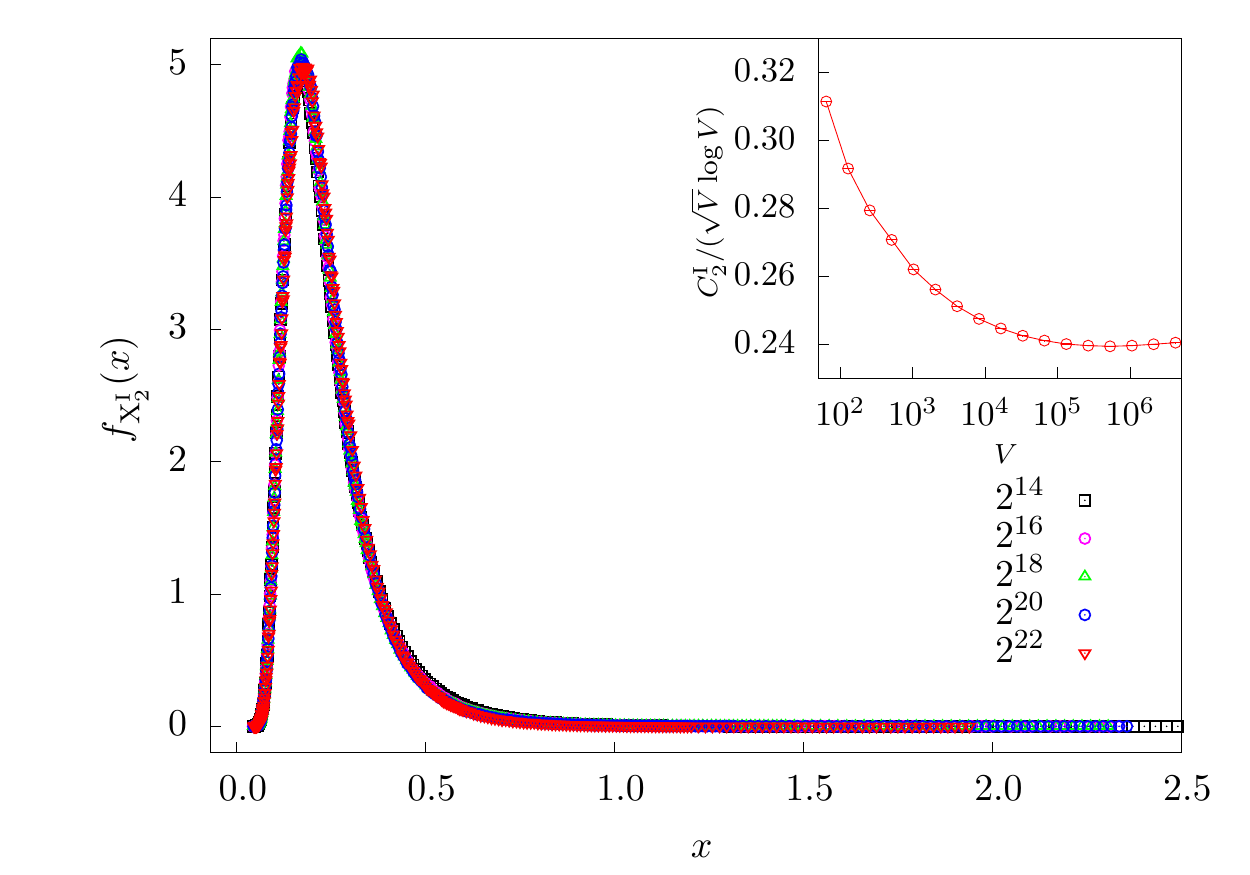}
    \caption{The probability distribution of $\scrC^{\rm I}_2$, the size of the second largest cluster in the Ising sector. Here $f_{X^{\rm I}_2}(x)$ is the probability density function of $X^{\rm I}_2 := \scrC^{\rm I}_2 / (\sqrt{V}\log V)$. The inset plots the average $C^{\rm I}_2/(\sqrt{V}\log V)$ versus $V$.}
    \label{Fig:distribution of C2 in Ising sector}
  \end{figure} 
  
    Since the largest-cluster size $C_1 \sim V^{2/3}$ for the percolation sector and $C_1 \sim V^{3/4}$
	for the Ising sector, the distinguishment of these two sectors can in principle be done in various ways. For instance, one set $\scrC_1 \leq 2V^{2/3}$ for percolation and otherwise for Ising, or $\scrC_1 \ge V^{3/4}$
	for Ising and otherwise for percolation. In practice, the situation is more complicated
	since the size distribution of $\scrC_1$ is continuous for finite system sizes. As can be seen in the top figure of Fig. 1, the probability distributions of $\scrC_1$ rescaled by $V^{3/4}$ collapse nicely when $\scrC_1/V^{3/4} > 1$. There is an intermediate region $2V^{2/3} < \scrC_1 < V^{3/4}$, in which the percolation and Ising behaviors are mixed. This intermediate region plays an irrelevant role in illustrating the physics in the percolation and Ising sectors while introduces much finite-size corrections. Thus, to demonstrate the scaling of other observables in the Ising sector, we choose to sample from configurations with $\scrC_1 \geq V^{3/4}$ to avoid finite-size corrections as much as possible.

We first study $\scrC^{\rm I}_2 := \scrC_2 | S_{\rm I}$, the size of the second largest cluster in the Ising sector. In Ref.~\cite{LuczakLuczak2008}, a limiting distribution for $\scrC_2$ rescaled by $\sqrt{V}\log V$ is presented. We thus expect this is also the right rescale factor for $\scrC^{\rm I}_2$. Define $X^{\rm I}_2 := \scrC^{\rm I}_2 / (\sqrt{V}\log V)$ and its probability density function $f_{X^{\rm I}_2}(x)$. In Fig.~\ref{Fig:distribution of C2 in Ising sector}, we plot the data for $f_{X^{\rm I}_2}(x)$ and observe excellent data collapse for various system sizes. We thus expect its average $C^{\rm I}_2 \sim \sqrt{V}\log V$. In the inset of Fig.~\ref{Fig:distribution of C2 in Ising sector}, we plot $C^{\rm I}_2 / (\sqrt{V}\log V)$ versus $V$, and it shows that this ratio tends to a constant as $V\rightarrow \infty$.

\begin{figure}
	\centering
	\includegraphics[scale=0.65]{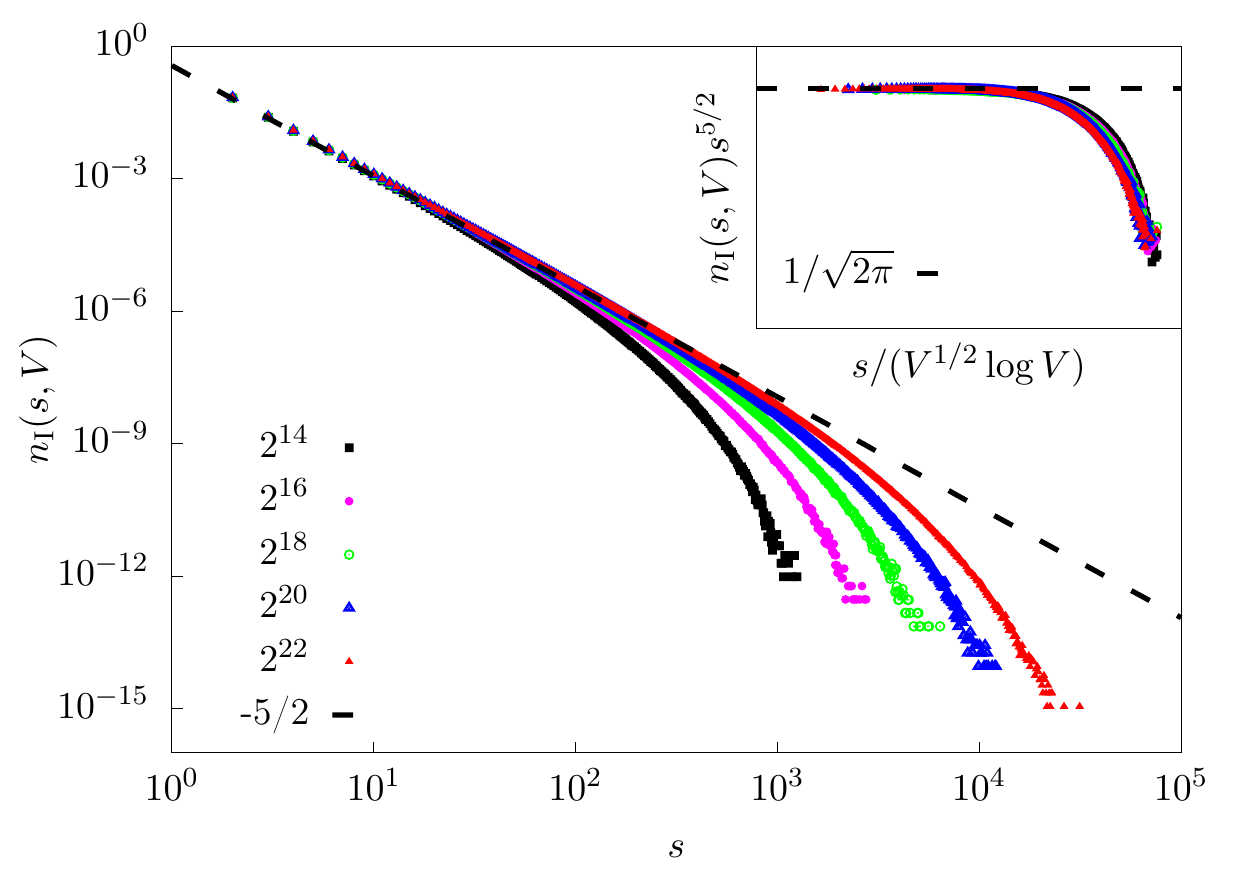}
	\caption{Log-log plot of the cluster-size distribution $n_{\rm I}(s,V)$ versus $s$ in the Ising sector. The slope of the black dashed line is $-5/2$. The inset shows the log-log plot of $n_{\rm I}(s,V)s^{5/2}$ versus $s/(\sqrt{V}\log V)$, which implies that the scaling function is consistent with $1/\sqrt{2\pi}$ when $s/(\sqrt{V}\log V) \ll 1$. }
	\label{Fig:cluster-size distribution in the Ising sector}
\end{figure}  

We then study the cluster-size distribution $n_{\rm I}(s,V)$ in the Ising sector. In Fig.~\ref{Fig:cluster-size distribution in the Ising sector}, we plot $n_{\rm I}(s,V)$ versus the cluster size $s$. Since in the Ising sector, the largest cluster dominates other clusters, we exclude the largest cluster in the plot of $n_{\rm I}(s,V)$ to avoid vast and meaningless discontinuity. Our data suggest $n_{\rm I}(s,V)\sim s^{-\tau}$ in the bulk region with $\tau = 5/2$, consistent with the Fisher exponent in the percolation case. In the inset of Fig.~\ref{Fig:cluster-size distribution in the Ising sector}, we plot $n_{\rm I}(s,V)s^{5/2}$ versus $s/(\sqrt{V}\log V)$, the well data collapse suggests the scaling $n_{\rm I}(s,V) \sim s^{-\tau}\tilde{n}_{\rm I}\left(\frac{s}{\sqrt{V}\log V}\right)$. Moreover, we again observed that the scaling function $\tilde{n}_{\rm I}(x)$ is consistent with the constant $1/\sqrt{2\pi}$ when $x\ll 1$. We note that, from the observed scaling formula of $n_{\rm I}(s,V)$, Eq.~\eqref{Eq:Fisher exponent} is violated regardless of whether $d_{\rm f}$ takes the fractal dimensions of $C_1$ or $C^{\rm I}_2$.

\subsection{The total sector}
Combined the results in the above three subsections, we make the following conjecture. The critical FK Ising model on ${\rm K}_V$ consists of a critical percolation sector and a critical Ising sector. The percolation sector decays asymptotically with a rate $V^{-1/12}$ and the Ising sector increases accordingly. The scaling of quantities in the FK ising model is a combination of their scaling in the percolation sector and the Ising sector, as conjectured in Eq.~\eqref{Eq:Conjectured quantities scaling}.

\begin{figure}
	\centering
	\includegraphics[scale=0.65]{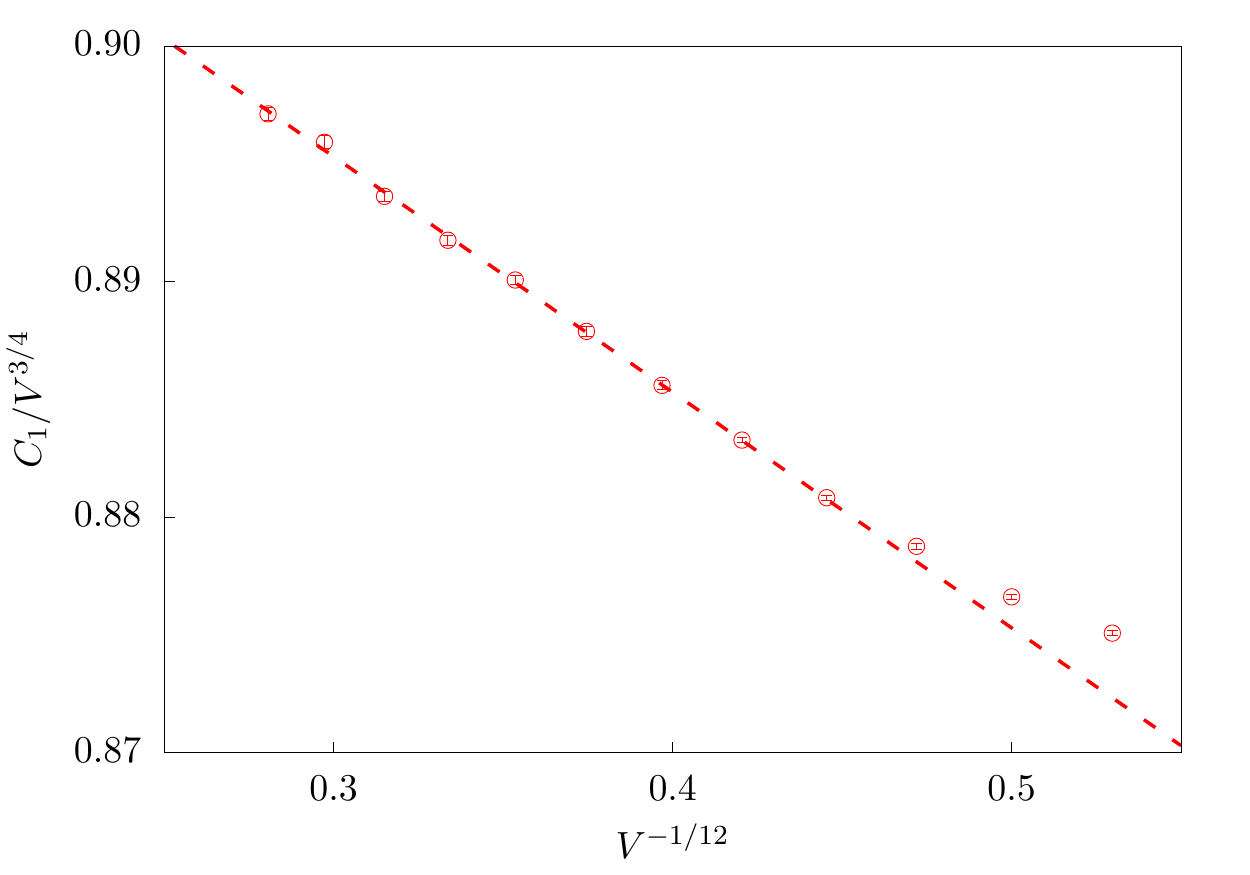}
	\caption{Plot to show the leading correction term of the size of the largest cluster $C_1$. A straight line with slope $1/12$ is to guide the eye.}
	\label{Fig:mean of C1}
\end{figure}

Since $C^{\rm P}_1 \sim V^{2/3}$ and $C^{\rm I}_1 \sim V^{3/4}$, by Eq.~\eqref{Eq:Conjectured quantities scaling} we expect that for the FK Ising model,
\begin{equation}
\label{Eq:conjecture for c1 scaling}
    C_1 = C^{\rm P}_1 \PP(S_{\rm P}) + C^{\rm I}_1 \PP(S_{\rm I}) \sim V^{3/4} - a_1V^{2/3} + a_2V^{7/12}\;,
\end{equation}
with some constants $a_1, a_2$. So, the effect of the percolation sector to the scaling of $C_1$ is subdominant. We then perform the least-square fits of the $C_1$ data to the ansatz $V^{3/4}(c_0 + c_1V^{y_1})$, and the fitting result shows $y_1 = 0.08(2)$, consistent with $1/12$. Fixing all three exponents in the RHS of Eq.~\eqref{Eq:conjecture for c1 scaling} to their expected values but leaving amplitudes free gives stable fits. In Fig.~\ref{Fig:mean of C1}, we plot $C_1/V^{3/4}$ versus $V^{-1/12}$, and the linearity shows the existence of the sub-dominant term $V^{2/3}$.

However, the effect of the percolation sector is not always subdominant. For the size of the second largest cluster, since $C^{\rm P}_2 \sim V^{2/3}$ and $C^{\rm I}_2 \sim \sqrt{V}\log V$ as observed in above sections, applying Eq.~\eqref{Eq:Conjectured quantities scaling} yields that
\begin{eqnarray}
\label{Eq:C2 scaling}
    C_2 & = & C^{\rm P}_2 \PP(S_{\rm P}) + C^{\rm I}_2 \PP(S_{\rm I})\nonumber \\
    &\sim &  V^{7/12} + a_3\sqrt{V}\log V - a_4 V^{5/12}\log V\;,
\end{eqnarray}
with some constants $a_3, a_4$. So, the effect of the percolation sector does dominate the scaling of $C_2$. Unfortunately, it is hard to numerically distinguish the three term in Eq.~\eqref{Eq:C2 scaling}, since their differences are quite small.

\begin{figure}
    \centering
    \includegraphics[scale=0.80]{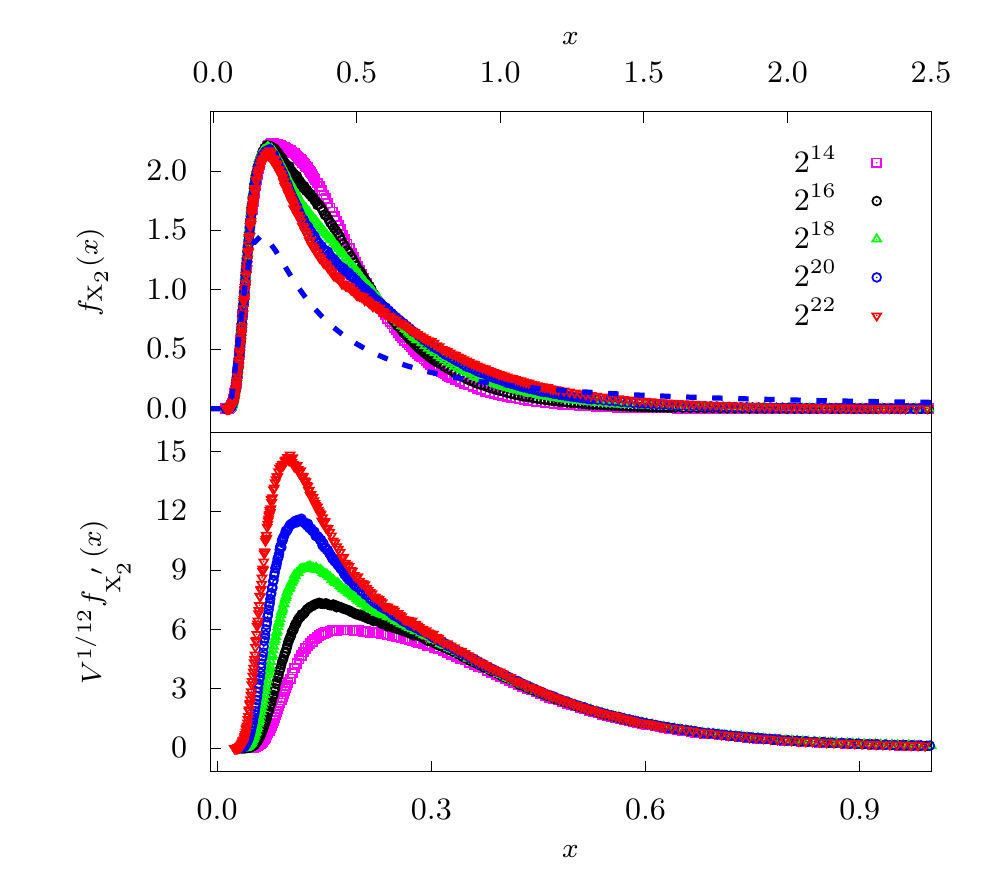}
    \caption{Plot of the distribution of $\scrC_2$, size of the second largest cluster. Here $f_{X_2}(x)$ (top) and $f_{X'_2}(x)$ (bottom) are respectively the probability density function of $X_2 = \scrC_2/(\sqrt{V}\log_{10} V)$ and $X'_2 = \scrC_2/V^{2/3}$. The curve in the top figure plots $h(x)$, shown in Eq.~\eqref{Eq:C2 distribution}, which is the limiting distribution of $X_2$ presented in Ref.~\cite{LuczakLuczak2008}.}
    \label{Fig:distribution of C2}
  \end{figure}

Moreover, although both the distribution of $\scrC^{\rm P}_2$ and $\scrC^{\rm I}_2$ show good data collapse, as shown in Fig.~\ref{Fig:distribution of C2 in percolation sector} and Fig.~\ref{Fig:distribution of C2 in Ising sector}, the distribution of $\scrC_2$ may fail to collapse due to the distinct scaling of $C^{\rm P}_2$ and $C^{\rm I}_2$. Define $X_2 = \scrC_2/(\sqrt{V}\log_{10} V)$ and its probability density function $f_{X_2}(x)$. We plot $f_{X_2}(x)$ in the top figure of Fig.~\ref{Fig:distribution of C2}. As shown, data for various systems collapse only when $x$ is small. We thus conjecture this collapsed region corresponds to the Ising sector and the rest (non-collapsed region) is contributed from the percolation sector. To confirm, we define $X'_2 = \scrC_2/V^{2/3}$ and its probability density function $f_{X'_2}(x)$. We plot $V^{1/12}f_{X'_2}(x)$ in the bottom figure of Fig.~\ref{Fig:distribution of C2}. Indeed, we observed a good data collapse when $x$ is approximately larger than $0.3$. Moreover, we note that, from the top figure of Fig.~\ref{Fig:distribution of C2}, it is unlikely that the distribution of $\scrC_2/(\sqrt{V}\log_{10} V)$ converges to the limiting distribution presented in Theorem 18 in Ref.~\cite{LuczakLuczak2008}, or it converges very slowly due to the strong finite-size corrections from percolation effects. A detailed study of the distribution of $\scrC_2$ is presented in Sec.~\ref{Appendix}.

\begin{figure}
    \centering
    \includegraphics[scale=0.7]{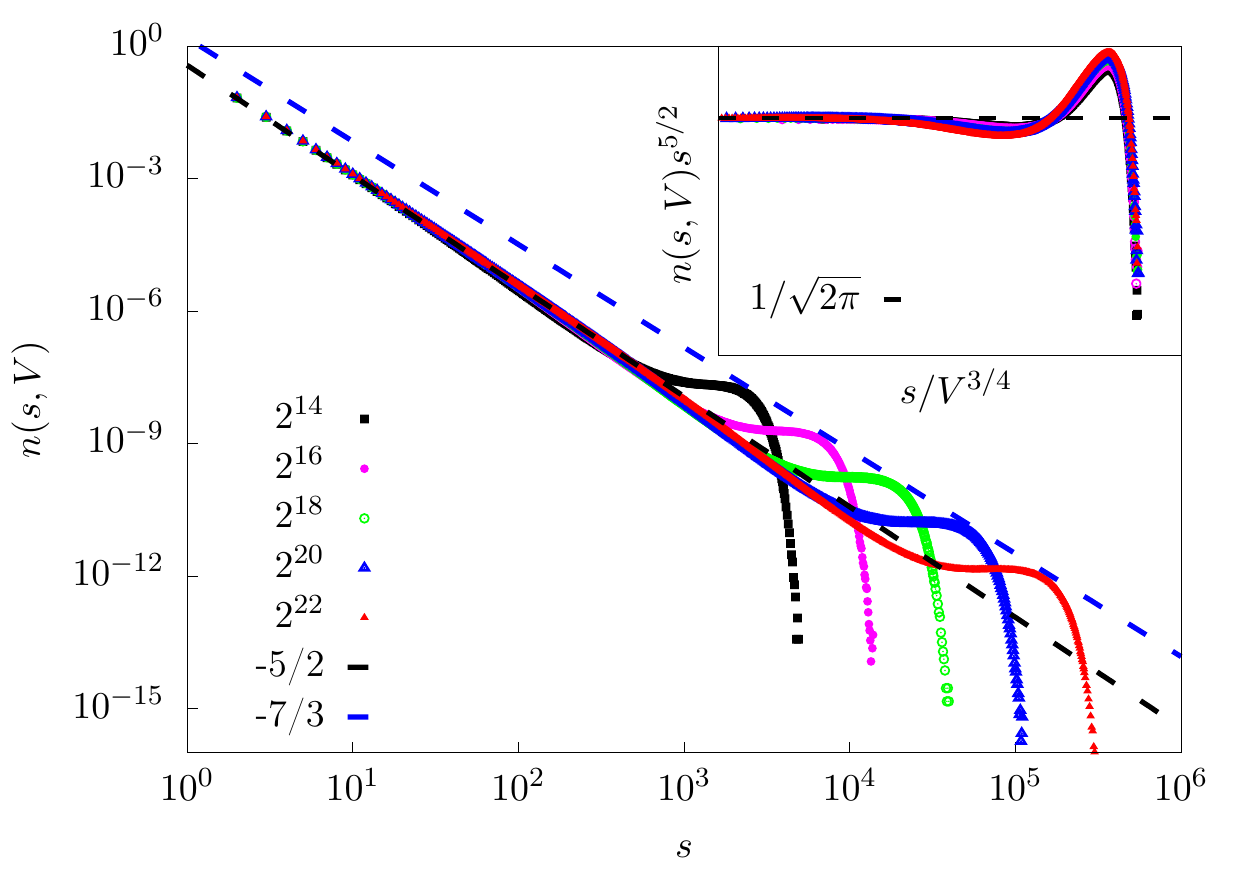}
    \caption{Log-log plot of $n(s,V)$ versus $s$ for various systems. The slopes of the two dashed lines are $5/2$ and $7/3$, respectively corresponding the Fisher exponent taking the percolation and Ising values. The inset plot $n(s,V)s^{5/2}$ versus $s/V^{3/4}$ in log-log scale, and the data suggest the scaling function is constant at $1/\sqrt{2\pi}$ when $s\ll V^{3/4}$.}
    \label{Fig:cluster-size distribution}
  \end{figure}

For the cluster-size distribution, since both $n_{\rm P}(s,V)$ and $n_{\rm I}(s,V)$ scale as $s^{-\tau}$ in the bulk region, see Fig.~\ref{Fig:cluster-size distribution in percolation sector} and Fig.~\ref{Fig:cluster-size distribution in the Ising sector}, we expect the same scaling for $n(s,V)$. In Fig.~\ref{Fig:cluster-size distribution} we plot $n(s,V)$ versus $s$ in log-log scale. Indeed one can observe that in the bulk region $n(s,V) \sim s^{-\tau}$ with $\tau = 5/2$, taking the percolation value. We also notice that, if the turning points of $n(s,V)$ data for each system size are connected, then such a line shows a slope $7/3$. Taking into Eq.~\eqref{Eq:Fisher exponent} gives $d_{\rm f} = 3/4$, which is the fractal dimension of $C_1$.

\subsection{Understanding the percolation scaling window from a RG-flow perspective}
\label{Sec:RG}
In this section, we propose an understanding to the two scaling windows of the FK Ising model, proved in Ref.~\cite{LuczakLuczak2008}, from the perspective of a renormalization-group(RG) flow. Consider the standard Ising model with a coupling constant $K$ on the complete graph ${\rm K}_V$. For each pair of spin variables $\sigma_i, \sigma_j$, define a bond random variable $b_{ij}\in\{0,1\}$ such that $\PP(b_{ij} = 1) = p \delta_{\sigma_i,\sigma_j}$ where $p$ is a free parameter. In words, we consider a bond percolation problem defined as follows. Given an Ising spin configuration via the Gibbs measure, one places bonds with probability $p$ between adjacent vertices with the same spin. For convenience, we further parameterize $p = 1 - e^{-2 K_{\rm P}}$. When $K_{\rm P} = K$, the bond percolation problem defined above is the FK Ising model. Such a generalized model was studied on lattices in Refs.~\cite{BloteKnopsNienhuis92,DengBlote04,qian2005percolation}.

\begin{figure}  
       \centering
      \includegraphics[scale=1.1]{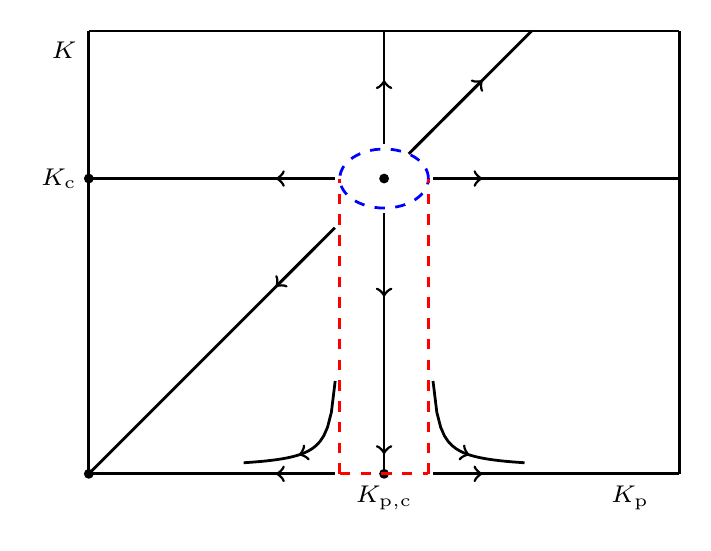}
    \caption{The $(K_{\rm P}, K)$ diagram and its associated RG flow. The line $K_{\rm P} = K$ corresponds the FK Ising model, and $(K_{{\rm P},{\rm c}}, K_{\rm c})$ is the critical point.}
           \label{Fig:RG-flow}
  \end{figure}

In Fig.~\ref{Fig:RG-flow}, a $(K_{\rm P}, K)$ diagram is shown to illustrate the RG flow. The whole line corresponding to $K_{\rm P} = 0$ is the percolation unstable fixed points since $p=0$ and no bond is placed on spin configurations. Along the vertical line from $(K_{{\rm P}, {\rm c}}, 0)$ to $(K_{{\rm P}, {\rm c}}, K_{\rm c})$, including the end points, there is no spontaneous breaking of the Ising symmetry, and the system is effectively of half plus spins and half minus spins. Thus, a simple analysis by ignoring the effect of fluctuations yields that, on average, the bond percolation is on two complete graphs, each of which has number of vertices as $V/2$, and thus the percolation threshold is $p_{\rm c}=2/V$, independent of the coupling strength $K$. Since this simple analysis holds true at the critical point $K_{\rm c}$, where the fluctuations are strongest, it is natural that $p_{\rm c}=2/V$ for the whole high-temperature phase $K <K_{\rm c}$. The percolation critical window is of size $O(V^{-1/3})$~\cite{Bollobas2001}, shown as the red column in Fig.~\ref{Fig:RG-flow}. We note that the flow out of the point $(K_{{\rm P},{\rm c}},K_{\rm c})$ along the line $K= K_{\rm c}$ is governed by the so-called red-bond exponent~\cite{Coniglio89}. Our simulations suggest that the red-bond exponent is consistent with $1/3$ on the complete graph. \par 
It was proved~\cite{LuczakLuczak2008} for the FK Ising model (along the line $K_{\rm P} = K$), the Ising critical window around the critical point $(K_{{\rm P},c},K_{\rm c})$ has size $O(V^{-1/2})$. This critical window is smaller than the one along the line $K = K_{\rm c}$, since the red-bond exponent is numerically found to be $1/3$, less than $1/2$. In Fig.~\ref{Fig:RG-flow}, we illustrate such an anisotropic critical window around $(K_{{\rm P},c},K_{\rm c})$ by a blue dashed ellipse. So, starting from the point $(K_{{\rm P},c}, K_{\rm c})$ and moving along the line $K_{\rm P} = K$ towards the origin, since the percolation critical window is asymptotically wider than the Ising one, the system will first be in Ising critical window and then enter the percolation scaling window.

\section{Discussion}
\label{Discussion}
For the FK Ising model on the complete graph, in addition to the percolation scaling window~\cite{LuczakLuczak2008}, we provide strong numerical evidence to show that, at the critical point, the FK Ising model possesses a percolation sector, which decays asymptotically with a rate $V^{-1/12}$. Quantities, such as the size of the largest and second largest clusters and the cluster-size distribution, conditioned in the percolation sector are shown to exhibit the same finite-size scaling as their counterparts in the critical percolation model on the complete graph. The effect of such a percolation sector is sub-dominant to the scaling of the largest cluster, but is dominant in the scaling of the second largest cluster. We also demonstrate that there is also a percolation scaling in the FK Ising model, which can be seen in the scaling of the cluster-size distribution. When clusters are much smaller than the largest cluster but larger than a constant, the scaling of the cluster-size distribution is governed by the Fisher exponent taking the percolation value $5/2$, rather than the Ising value $7/3$. It would be interesting if some of these numerical observations can be established rigorously, for example, the exponent $1/12$ which is the rate the percolation sector decays with.

Field theory predicts that, on a lattice, the upper critical dimension $d_{\rm c} = 4$ for the FK Ising model and $d_{\rm c} = 6$ for percolation. In terms of finite-size scaling, one would expect that these models on hypercubic lattices with dimension $d > d_{\rm c}$ and periodic boundary conditions exhibit their complete-graph asymptotics. However, for the FK Ising model with $d=5$, as numerically studied in Ref.~\cite{FangGrimmZhouDeng20}, the above statement is \emph{only} true for the size of the largest cluster. The scaling of other clusters, such as the fractal dimension and cluster-size distribution, still follows the prediction from Gaussian fixed point, which predicts the thermal and magnetic exponents $(y_{\rm t}, y_{\rm h}) = (2, 1 + d/2)$. For example, for the $d=5$ case, the cluster-size distribution of clusters other than the largest one exhibits the Gaussian fixed point scaling, instead of the complete-graph percolation scaling which is observed in this paper for the FK Ising model on the complete graph. The potential reason could be that $5$ is still below the upper critical dimension of percolation. Above $d_{\rm c} = 6$, since the thermal and magnetic exponents for the complete-graph percolation are $(1/3, 2/3)$, it predicts that $(y_{\rm t}, y_{\rm h}) = (d/3, 2d/3)$ for percolation on lattices with periodic boundary conditions.  
So, for $d>6$, the exponents $(d/3, 2d/3)$ dominate $(2, 1 + d/2)$, and we thus conjecture that the FK Ising model on the hypercubic lattice with $d>6$ completely follows complete-graph asymptotics. In other words, we conjecture that, for the finite-size scaling of the FK Ising model on lattices, there are two special dimensions $4$ and $6$. For $d>4$, only the largest cluster exhibits the complete-graph asymptotics, but for $d>6$, all clusters follow the complete-graph asymptotics.

 \section{Acknowledgments}
Y.D. acknowledges support by the National Natural Science Foundation of China under Grant No. 11625522 and by the National Key R\&D Program of China under Grant No. 2018YFA0306501. Z.Z. and S.F. thank the Research Support Scheme from ACEMS for providing financial support to hospitalize S. F in Monash University, during which this work was written. We also thank the Supercomputing Center of University of Science and Technology of China for the computer time. We thank Jens Grimm and Timothy Garoni for valuable discussions. 

\appendix
\section{The distribution of $\scrC_2$}
\label{Appendix}
We discuss in this section the size distribution of the second largest cluster $\scrC_2$. Recall that $X_2 = \scrC_2/(\sqrt{V}\log_{10} V)$. Theorem 18 in Ref.~\cite{LuczakLuczak2008} presents the limiting theorem for $X_2$, by which one can derive the probability density function of $X_2$ in the $V\rightarrow \infty$ limit is
\begin{equation}
\label{Eq:C2 distribution}
h(x) = \frac{x^{-3/2}\exp(-\frac{1}{48x^2})}{\int_{0}^\infty x^{-3/2}\exp(-\frac{1}{48x^2}) \dx}
\end{equation}
However, from the top figure in Fig.~\ref{Fig:distribution of C2}, it shows that the observed distribution of $X_2$ is less likely to converge to $h(x)$, especially in large $x$ region.

\begin{figure}  
       \centering
      \includegraphics[scale=0.7]{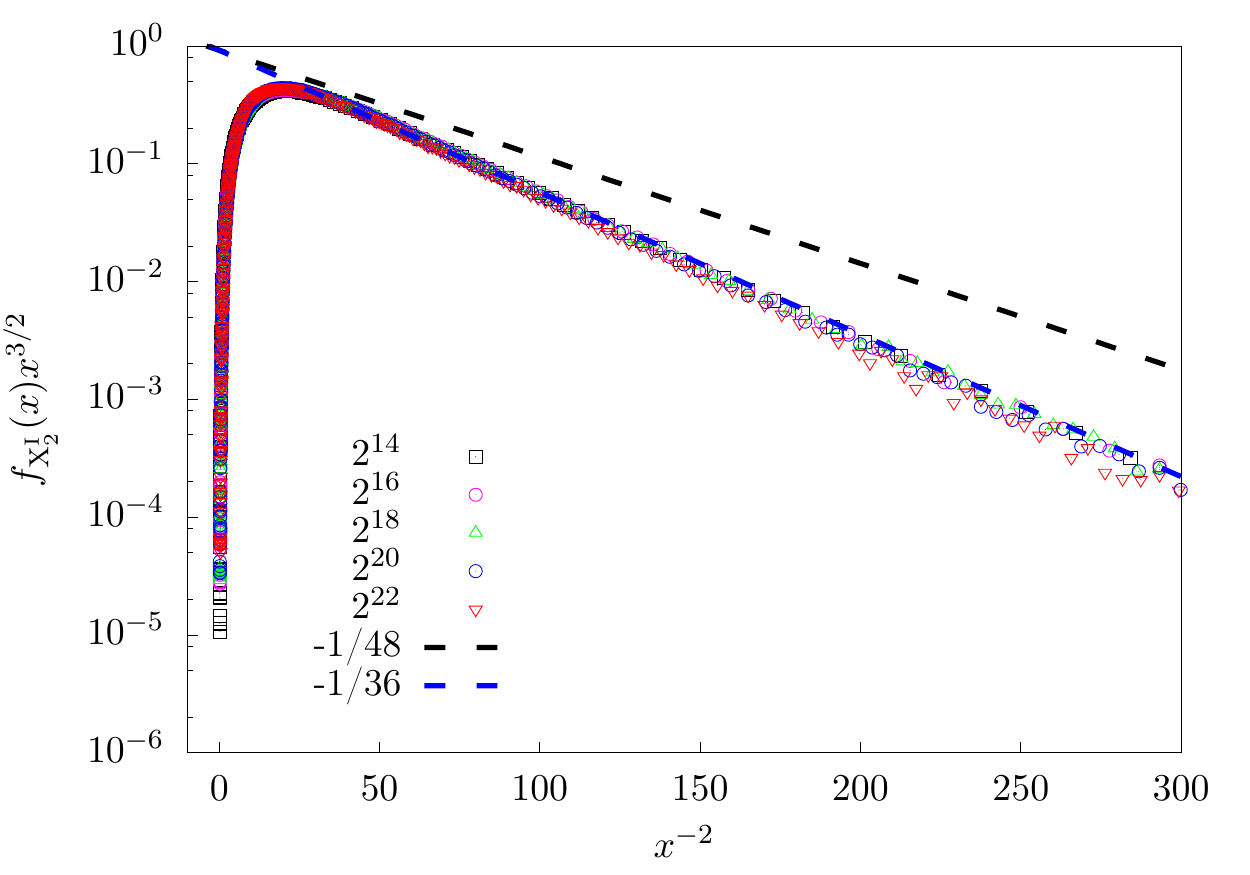}
    \caption{Plot to show the distribution of $\scrC^{\rm I}_2$, the size of the second largest cluster in the Ising sector. Here $X^{\rm I}_2 := \scrC^{\rm I}_2/(\sqrt{V}\log_{10} V)$ and $f_{X^{\rm I}_2}(x)$ is its probability density function. It can be seen that, for $x<x_0$ with $x_0\approx 0.2$, data from various systems are observed to collapse onto a straight line with slope $-1/36$.}
           \label{Fig:distribution of C2 small x}
  \end{figure}  

Since our data provide strong evidence for the existence of a percolation sector, we first discuss the impact of such a sector to the distribution of $\scrC_2$. For any $k>0$, one can write
\begin{equation}
\PP(\scrC_2 \leq k) = \PP(\scrC_2\leq k | S_{\rm P}) \PP(S_{\rm P}) +  \PP(\scrC_2\leq k|S_{\rm I}) \PP(S_{\rm I}) \;. \nonumber
\end{equation}
But, it is observed that $\PP(S_{\rm P}) \sim V^{-1/12}$ which vanishes as $V$ tends to infinity. This means that the percolation sector has zero impact  to the limiting distribution of $\scrC_2$, and therefore $\scrC_2$ has the same limiting distribution as $\scrC^{\rm I}_2$, the size of the second largest cluster conditioned in the Ising sector.

We then focus on the distribution of $\scrC^{\rm I}_2$. Recall that $X^{\rm I}_2 := \scrC^{\rm I}_2/(\sqrt{V}\log_{10} V)$. Fig.~\ref{Fig:distribution of C2 in Ising sector} shows quite well data collapse for the distribution of $X^{\rm I}_2$, however, it is still less likely to converge to $h(x)$ as $V$ tends to infinity. To study the distribution of $X^{\rm I}_2$, we first plot $\ln \left[ f_{X^{\rm I}_2}(x) x^{3/2}\right]$ versus $x^{-2}$ in Fig.~\ref{Fig:distribution of C2 small x}. Clearly, when $x$ is smaller than a constant $x_0$ ($x_0\approx 0.2$, and $x<x_0$ corresponds to the increasing region in Fig.~\ref{Fig:distribution of C2 in Ising sector}), our data collapse to a straight line with slope $1/36$. So, we conjecture, in the region $x < x_0$, the probability density function of $X^{\rm I}_2$ converges to
\begin{equation}
\label{Eq:f_X2(x) for small x}
f^{\infty}_{X_2}(x) \propto x^{-3/2}\exp\left(-\frac{1}{36 x^2}\right)\;.
\end{equation} 
We note that the function above differs with $h(x)$ defined in Eq.~\eqref{Eq:C2 distribution}, that is, for the constant in front of $x^{-2}$ inside the exponential, our data is in favor of $1/36$, instead of $1/48$. Possibly, such a discrepancy is due to some strong finite-size corrections, which are hard to be detected using the system sizes achieved in our simulations. 

\section{Critical behaviour of the reduced susceptibility}
The susceptibility of the FK Ising model on the complete graph is defined as $\chi = \sum_{i}\left\langle {\scrC^2_i}\right\rangle/V$, where $\scrC_i$ is the size of the $i$-th largest cluster. Under the spin representation, $\chi$ can be written as the fluctuation of the magnetization, and it can be analytically shown that $\chi \sim V^{1/2}$ at the critical point \cite{luijten1997interaction}. On the hypercubic lattices with periodic boundary conditions and above the upper critical dimension ($\dc =4$), it is believed that the scaling of $\chi$ follows the complete-graph asymptotics by setting $V=L^d$, i.e., $\chi \sim L^{d/2}$, which has been numerically confirmed at $d=5$~\cite{WittmannYoung2014,GrimmElciZhouGaroniDeng2017,ZhouGrimmFangDengGaroni2018}. 

The reduced susceptibility $\chi^\prime$ is defined similarly but excluding the largest cluster in the summation, i.e., $\chi^\prime = \sum_{i\neq 1} \left\langle {\scrC^2_i} \right\rangle/V$. Compared with $\chi$, the finite-size scaling of $\chi^\prime$ turns out to be more subtle. The critical behaviour of $\chi^\prime$ at and near the critical point was studied on five-dimensional torus in Ref.~\cite{FangGrimmZhouDeng20}. This section presents our numerical results for $\chi^\prime$ on the complete graph.

\begin{figure}[H]
\centering
\includegraphics[scale=0.6]{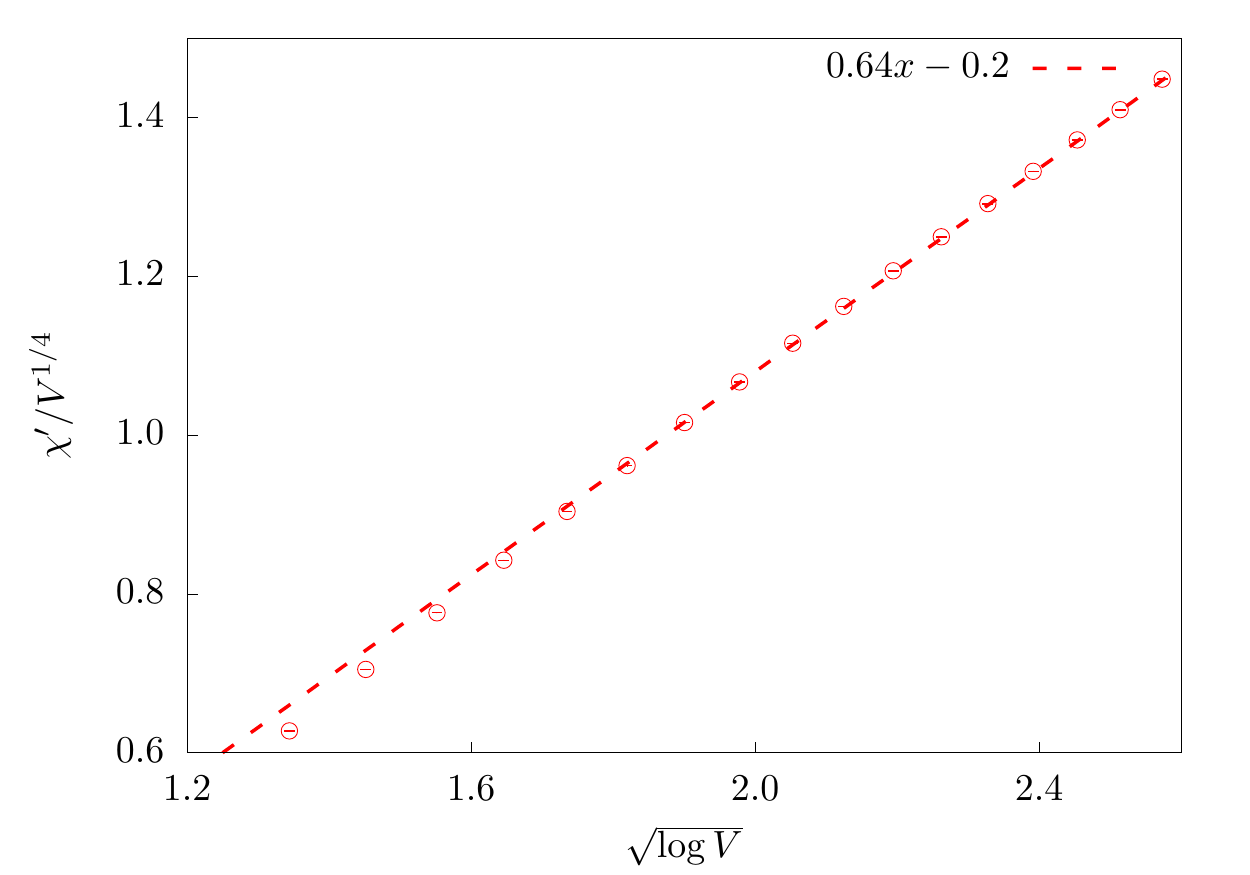}
\caption{Finite-size scaling of the reduced susceptibility $\chi'$ at the critical point $p_c$.}
\label{Fig:chi_c}
\end{figure}

We start at the critical point $\pc$. Our results in previous sections suggest that the cluster number density $n(s,V) \sim s^{-5/2}$ when $s=O(\sqrt{V}\log V)$, from which $\chi^\prime$ can be approximated as
\begin{equation}
\chi^\prime \approx \int_{1}^{\sqrt{V}\log V} s^2n(s,V) \ds \sim V^{1/4}\sqrt{\log V}
\end{equation}
In Fig.~\ref{Fig:chi_c}, we plot $\chi^\prime/V^{1/4}$ versus $\sqrt{\log V}$ and observe that the large-$V$ data collapse onto a straight line with non-zero intercept. We then perform the least-square fits of $\chi^\prime$ to the ansatz $\chi' = V^{y_{\chi'}} (a_0 (\log V)^{\hat{y}_{\chi'}}+ a_1)$. Leaving both $y_{\chi'}$ and $\hat{y}_{\chi'}$ free cannot produce stable fitting. As shown in Table~\ref{Tab:chi}, when fix $\hat{y}_{\chi'} = 1/2$, our fits estimate $y_{\chi'} =0.251(1)$; when fix $y_{\chi'} =1/4 $, we can estimate $\hat{y}_{\chi'}  = 0.53(6)$. Both suggest $\chi^\prime \sim V^{1/4}\sqrt{\log V}$.

\begin{table}[H]
 \centering
 \begin{tabular}{lccccc}
  \hline 
  $V_{\rm min}$   &$a_0$  &$a_1$  &$y_{\chi'}$  &$\hat{y}_{\chi'}$  & ${\rm chi}^2/{\rm DF}$  \\
  \hline 
  $2^{15}$ &$0.64(1)$  &$-0.20(2)$    &$0.2498(5)$ &$1/2$  &$4.0/5$\\ 
  $2^{16}$ &$0.62(2)$  &$-0.16(3)$    &$0.2508(9)$ &$1/2$  &$2.0/4$\\ 
  \hline 
  $2^{15}$ &$0.66(4)$  &$-0.22(5)$    &$1/4$  &$0.49(2)$  &$4.0/5$\\ 
  $2^{16}$ &$0.58(5)$  &$-0.11(6)$    &$1/4$  &$0.53(2)$  &$2.0/4$\\ 
  \hline
 \end{tabular} 
\caption{Fitting results for $\chi'$ at $\pc$. No error bars are presented if we chose to fix a specific number for the corresponding parameter. In the last column, ${\rm chi}^2$ and DF stand for the residual and degree of freedom in the fitting, respectively.}
\label{Tab:chi}
\end{table}
 
\begin{figure*}[ht!]
 	\centering
 	\includegraphics[scale=0.7]{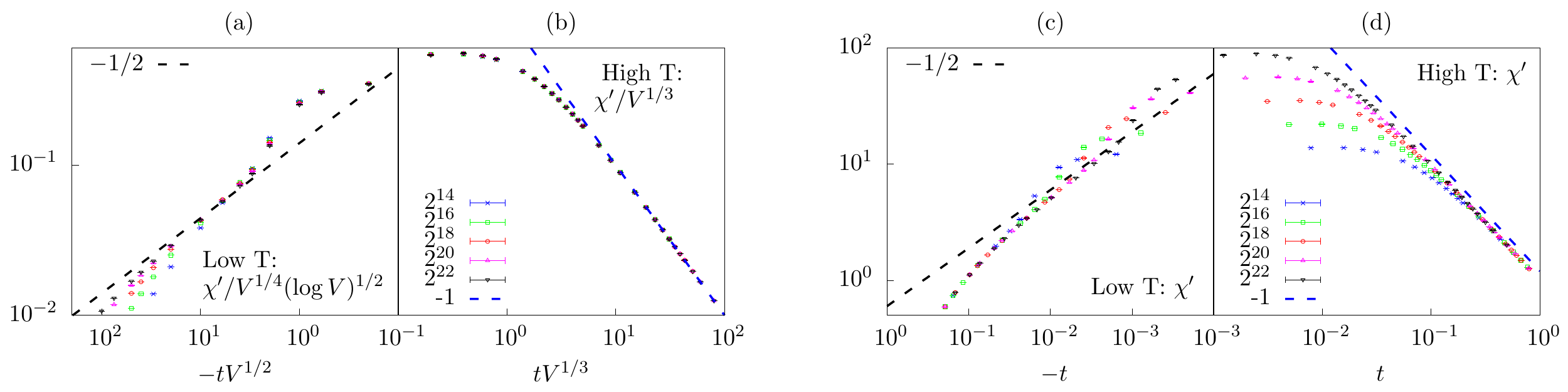}
 	\caption{The finite-size scaling of the reduced susceptibility $\chi^\prime$ in low-T $(a)$ and high-T $(b)$ critical windows, and the thermodynamic-limit behaviour of $\chi^\prime$ in low-T $(c)$ and high-T $(d)$ regions. Same range of vertical axis applies to $(a)$ and $(b)$, and to $(c)$ and $(d)$.}
 	\label{Fig:chiprime}
\end{figure*}

Finally we study $\chi'$ near the critical point $\pc$. Define $t = (\pc - p)/\pc$. We first consider the high-temperature(T) case ($t>0$). In Fig.~\ref{Fig:chiprime}(b), we plot $\chi^\prime(t,V)/V^{1/3}$ versus $tV^{1/3}$ for various systems, and the excellent data collapse suggests the finite-size scaling $\chi^\prime(t,V) \sim V^{2\yh - 1} \tilde{\chi}^\prime_{\rm h}(tV^{\yt})$, where $\tilde{\chi}^\prime_{\rm h}(\cdot)$ is the scaling function and the thermal and magnetic critical exponents $(\yt, \yh) = (1/3, 2/3)$, same as those for percolation on the complete graph. To recover the thermodynamic-limit behaviour $\chi^\prime(t,\infty) \sim t^{-\gamma_{\rm h}}$, one expects $\tilde{\chi}^\prime_{\rm h}(x) \sim x^{-\gamma_{\rm h}}$ and $\gamma_{\rm h} = (2\yh -1)/\yt = 1$. This is numerically confirmed in Fig.~\ref{Fig:chiprime}(b) and (d).

In the low-T region, we plot in Fig.~\ref{Fig:chiprime}(a) $\chi^\prime(t,V)/\sqrt{V^{1/2} \log V}$ versus $tV^{1/2}$ for various systems, and observe good data collapse when $t$ is close to zero. This suggests the finite-size scaling $\chi^\prime(t,V) \sim V^{1/4}\sqrt{\log V} \tilde{\chi}^\prime_{\rm l}(tV^{1/2})$. To recover the thermodynamic-limit behaviour, one would expect $\tilde{\chi}^\prime_{\rm l}(x) \sim x^{-1/2}$ such that $\chi^\prime(t,\infty) \sim t^{-1/2}$. However, as shown in Fig.~\ref{Fig:chiprime}(a) and (c), in low-T region $\chi^\prime$ become off-critical quite quickly such that the power-law scaling can hardly be observed.

\bibliographystyle{apsrev4-1}

\begin{thebibliography}{21}%
\makeatletter
\providecommand \@ifxundefined [1]{%
 \@ifx{#1\undefined}
}%
\providecommand \@ifnum [1]{%
 \ifnum #1\expandafter \@firstoftwo
 \else \expandafter \@secondoftwo
 \fi
}%
\providecommand \@ifx [1]{%
 \ifx #1\expandafter \@firstoftwo
 \else \expandafter \@secondoftwo
 \fi
}%
\providecommand \natexlab [1]{#1}%
\providecommand \enquote  [1]{``#1''}%
\providecommand \bibnamefont  [1]{#1}%
\providecommand \bibfnamefont [1]{#1}%
\providecommand \citenamefont [1]{#1}%
\providecommand \href@noop [0]{\@secondoftwo}%
\providecommand \href [0]{\begingroup \@sanitize@url \@href}%
\providecommand \@href[1]{\@@startlink{#1}\@@href}%
\providecommand \@@href[1]{\endgroup#1\@@endlink}%
\providecommand \@sanitize@url [0]{\catcode `\\12\catcode `\$12\catcode
  `\&12\catcode `\#12\catcode `\^12\catcode `\_12\catcode `\%12\relax}%
\providecommand \@@startlink[1]{}%
\providecommand \@@endlink[0]{}%
\providecommand \url  [0]{\begingroup\@sanitize@url \@url }%
\providecommand \@url [1]{\endgroup\@href {#1}{\urlprefix }}%
\providecommand \urlprefix  [0]{URL }%
\providecommand \Eprint [0]{\href }%
\providecommand \doibase [0]{http://dx.doi.org/}%
\providecommand \selectlanguage [0]{\@gobble}%
\providecommand \bibinfo  [0]{\@secondoftwo}%
\providecommand \bibfield  [0]{\@secondoftwo}%
\providecommand \translation [1]{[#1]}%
\providecommand \BibitemOpen [0]{}%
\providecommand \bibitemStop [0]{}%
\providecommand \bibitemNoStop [0]{.\EOS\space}%
\providecommand \EOS [0]{\spacefactor3000\relax}%
\providecommand \BibitemShut  [1]{\csname bibitem#1\endcsname}%
\let\auto@bib@innerbib\@empty
\bibitem [{\citenamefont {Grimmett}(2006)}]{Grimmett2006}%
  \BibitemOpen
  \bibfield  {author} {\bibinfo {author} {\bibfnamefont {G.}~\bibnamefont
  {Grimmett}},\ }\href {https://books.google.com.au/books?id=UfvxyLIMalgC}
  {\emph {\bibinfo {title} {The Random-Cluster Model}}},\ Grundlehren der
  mathematischen Wissenschaften\ (\bibinfo  {publisher} {Springer Berlin
  Heidelberg},\ \bibinfo {year} {2006})\BibitemShut {NoStop}%
\bibitem [{\citenamefont {Kasteleyn}\ and\ \citenamefont
  {Fortuin}(1969)}]{kasteleyn1969phase}%
  \BibitemOpen
  \bibfield  {author} {\bibinfo {author} {\bibfnamefont {P.W.}~\bibnamefont
  {Kasteleyn}}\ and\ \bibinfo {author} {\bibfnamefont {C.M.}~\bibnamefont
  {Fortuin}},\ }\href@noop {} {\bibfield  {journal} {\bibinfo  {journal}
  {J. Phys. Soc. Jpn. Suppl.}\ }\textbf {\bibinfo
  {volume} {26}},\ \bibinfo {pages} {11} (\bibinfo {year} {1969})}\BibitemShut
  {NoStop}%
\bibitem [{\citenamefont {Bollob{\'a}s}\ \emph {et~al.}(1996)\citenamefont
  {Bollob{\'a}s}, \citenamefont {Grimmett},\ and\ \citenamefont
  {Janson}}]{BollobasGrimmettJanson1996}%
  \BibitemOpen
  \bibfield  {author} {\bibinfo {author} {\bibfnamefont {B.}~\bibnamefont
  {Bollob{\'a}s}}, \bibinfo {author} {\bibfnamefont {G.}~\bibnamefont
  {Grimmett}}, \ and\ \bibinfo {author} {\bibfnamefont {S.}~\bibnamefont
  {Janson}},\ }\href@noop {} {\bibfield  {journal} {\bibinfo  {journal}
  {Probability Theory and Related Fields}\ }\textbf {\bibinfo {volume} {104}},\
  \bibinfo {pages} {283} (\bibinfo {year} {1996})}\BibitemShut {NoStop}%
\bibitem [{\citenamefont {Luczak}\ and\ \citenamefont
  {\L~uczak}(2006)}]{LuczakLuczak2008}%
  \BibitemOpen
  \bibfield  {author} {\bibinfo {author} {\bibfnamefont {M.}~\bibnamefont
  {Luczak}}\ and\ \bibinfo {author} {\bibfnamefont {T.}~\bibnamefont
  {\L~uczak}},\ }\href@noop {} {\bibfield  {journal} {\bibinfo  {journal}
  {Random Structures \& Algorithms}\ }\textbf {\bibinfo {volume} {28}},\
  \bibinfo {pages} {215} (\bibinfo {year} {2006})}\BibitemShut {NoStop}%
\bibitem [{\citenamefont {Lundow}\ and\ \citenamefont
  {Markstr{\"o}m}(2015)}]{lundow2015complete}%
  \BibitemOpen
  \bibfield  {author} {\bibinfo {author} {\bibfnamefont {P.-H.}\ \bibnamefont
  {Lundow}}\ and\ \bibinfo {author} {\bibfnamefont {K.}~\bibnamefont
  {Markstr{\"o}m}},\ }\href@noop {} {\bibfield  {journal} {\bibinfo  {journal}
  {Physical Review E}\ }\textbf {\bibinfo {volume} {91}},\ \bibinfo {pages}
  {022112} (\bibinfo {year} {2015})}\BibitemShut {NoStop}%
\bibitem [{\citenamefont {Huang}\ \emph {et~al.}(2018)\citenamefont {Huang},
  \citenamefont {Hou}, \citenamefont {Wang}, \citenamefont {Ziff},\ and\
  \citenamefont {Deng}}]{huang2018critical}%
  \BibitemOpen
  \bibfield  {author} {\bibinfo {author} {\bibfnamefont {W.}~\bibnamefont
  {Huang}}, \bibinfo {author} {\bibfnamefont {P.}~\bibnamefont {Hou}}, \bibinfo
  {author} {\bibfnamefont {J.}~\bibnamefont {Wang}}, \bibinfo {author}
  {\bibfnamefont {R.~M.}\ \bibnamefont {Ziff}}, \ and\ \bibinfo {author}
  {\bibfnamefont {Y.}~\bibnamefont {Deng}},\ }\href@noop {} {\bibfield
  {journal} {\bibinfo  {journal} {Physical Review E}\ }\textbf {\bibinfo
  {volume} {97}},\ \bibinfo {pages} {022107} (\bibinfo {year}
  {2018})}\BibitemShut {NoStop}%
\bibitem [{\citenamefont {Hou}\ \emph {et~al.}(2019)\citenamefont {Hou},
  \citenamefont {Fang}, \citenamefont {Wang}, \citenamefont {Hu},\ and\
  \citenamefont {Deng}}]{hou2019geometric}%
  \BibitemOpen
  \bibfield  {author} {\bibinfo {author} {\bibfnamefont {P.}~\bibnamefont
  {Hou}}, \bibinfo {author} {\bibfnamefont {S.}~\bibnamefont {Fang}}, \bibinfo
  {author} {\bibfnamefont {J.}~\bibnamefont {Wang}}, \bibinfo {author}
  {\bibfnamefont {H.}~\bibnamefont {Hu}}, \ and\ \bibinfo {author}
  {\bibfnamefont {Y.}~\bibnamefont {Deng}},\ }\href@noop {} {\bibfield
  {journal} {\bibinfo  {journal} {Physical Review E}\ }\textbf {\bibinfo
  {volume} {99}},\ \bibinfo {pages} {042150} (\bibinfo {year}
  {2019})}\BibitemShut {NoStop}%
\bibitem [{\citenamefont {Ben-Naim}\ and\ \citenamefont
  {Krapivsky}(2005)}]{ben2005kinetic}%
  \BibitemOpen
  \bibfield  {author} {\bibinfo {author} {\bibfnamefont {E.}~\bibnamefont
  {Ben-Naim}}\ and\ \bibinfo {author} {\bibfnamefont {P.}~\bibnamefont
  {Krapivsky}},\ }\href@noop {} {\bibfield  {journal} {\bibinfo  {journal}
  {Physical Review E}\ }\textbf {\bibinfo {volume} {71}},\ \bibinfo {pages}
  {026129} (\bibinfo {year} {2005})}\BibitemShut {NoStop}%
\bibitem [{\citenamefont {Swendsen}\ and\ \citenamefont
  {Wang}(1987)}]{swendsen1987nonuniversal}%
  \BibitemOpen
  \bibfield  {author} {\bibinfo {author} {\bibfnamefont {R.~H.}\ \bibnamefont
  {Swendsen}}\ and\ \bibinfo {author} {\bibfnamefont {J.-S.}\ \bibnamefont
  {Wang}},\ }\href@noop {} {\bibfield  {journal} {\bibinfo  {journal} {Physical
  review letters}\ }\textbf {\bibinfo {volume} {58}},\ \bibinfo {pages} {86}
  (\bibinfo {year} {1987})}\BibitemShut {NoStop}%
\bibitem [{\citenamefont {Wolff}(1989{\natexlab{a}})}]{wolff1989collective}%
  \BibitemOpen
  \bibfield  {author} {\bibinfo {author} {\bibfnamefont {U.}~\bibnamefont
  {Wolff}},\ }\href@noop {} {\bibfield  {journal} {\bibinfo  {journal}
  {Physical Review Letters}\ }\textbf {\bibinfo {volume} {62}},\ \bibinfo
  {pages} {361} (\bibinfo {year} {1989}{\natexlab{a}})}\BibitemShut {NoStop}%
\bibitem [{\citenamefont {Wolff}(1989{\natexlab{b}})}]{wolff1989comparison}%
  \BibitemOpen
  \bibfield  {author} {\bibinfo {author} {\bibfnamefont {U.}~\bibnamefont
  {Wolff}},\ }\href@noop {} {\bibfield  {journal} {\bibinfo  {journal} {Physics
  Letters B}\ }\textbf {\bibinfo {volume} {228}},\ \bibinfo {pages} {379}
  (\bibinfo {year} {1989}{\natexlab{b}})}\BibitemShut {NoStop}%
\bibitem [{\citenamefont {Nachmias}\ and\ \citenamefont
  {Peres}(2008)}]{nachmias2008critical}%
  \BibitemOpen
  \bibfield  {author} {\bibinfo {author} {\bibfnamefont {A.}~\bibnamefont
  {Nachmias}}\ and\ \bibinfo {author} {\bibfnamefont {Y.}~\bibnamefont
  {Peres}},\ }\href@noop {} {\bibfield  {journal} {\bibinfo  {journal} {The
  Annals of Probability}\ }\textbf {\bibinfo {volume} {36}},\ \bibinfo {pages}
  {1267} (\bibinfo {year} {2008})}\BibitemShut {NoStop}%
\bibitem [{\citenamefont {Bl\"ote}\ \emph {et~al.}(1992)\citenamefont
  {Bl\"ote}, \citenamefont {Knops},\ and\ \citenamefont
  {Nienhuis}}]{BloteKnopsNienhuis92}%
  \BibitemOpen
  \bibfield  {author} {\bibinfo {author} {\bibfnamefont {H.~W.~J.}\
  \bibnamefont {Bl\"ote}}, \bibinfo {author} {\bibfnamefont {Y.~M.~M.}\
  \bibnamefont {Knops}}, \ and\ \bibinfo {author} {\bibfnamefont
  {B.}~\bibnamefont {Nienhuis}},\ }\href {\doibase 10.1103/PhysRevLett.68.3440}
  {\bibfield  {journal} {\bibinfo  {journal} {Phys. Rev. Lett.}\ }\textbf
  {\bibinfo {volume} {68}},\ \bibinfo {pages} {3440} (\bibinfo {year}
  {1992})}\BibitemShut {NoStop}%
\bibitem [{\citenamefont {Deng}\ and\ \citenamefont
  {Bl\"ote}(2004)}]{DengBlote04}%
  \BibitemOpen
  \bibfield  {author} {\bibinfo {author} {\bibfnamefont {Y.}~\bibnamefont
  {Deng}}\ and\ \bibinfo {author} {\bibfnamefont {H.~W.~J.}\ \bibnamefont
  {Bl\"ote}},\ }\href {\doibase 10.1103/PhysRevE.70.056132} {\bibfield
  {journal} {\bibinfo  {journal} {Phys. Rev. E}\ }\textbf {\bibinfo {volume}
  {70}},\ \bibinfo {pages} {056132} (\bibinfo {year} {2004})}\BibitemShut
  {NoStop}%
\bibitem [{\citenamefont {Qian}\ \emph {et~al.}(2005)\citenamefont {Qian},
  \citenamefont {Deng},\ and\ \citenamefont {Bl{\"o}te}}]{qian2005percolation}%
  \BibitemOpen
  \bibfield  {author} {\bibinfo {author} {\bibfnamefont {X.}~\bibnamefont
  {Qian}}, \bibinfo {author} {\bibfnamefont {Y.}~\bibnamefont {Deng}}, \ and\
  \bibinfo {author} {\bibfnamefont {H.~W.}\ \bibnamefont {Bl{\"o}te}},\
  }\href@noop {} {\bibfield  {journal} {\bibinfo  {journal} {Physical Review
  B}\ }\textbf {\bibinfo {volume} {71}},\ \bibinfo {pages} {144303} (\bibinfo
  {year} {2005})}\BibitemShut {NoStop}%
\bibitem [{\citenamefont {Bollob{\'a}s}(2001)}]{Bollobas2001}%
  \BibitemOpen
  \bibfield  {author} {\bibinfo {author} {\bibfnamefont {B.}~\bibnamefont
  {Bollob{\'a}s}},\ }\href@noop {} {\emph {\bibinfo {title} {Random Graphs}}},\
  Cambridge Studies in Advanced Mathematics\ (\bibinfo  {publisher} {Cambridge
  University Press, New York},\ \bibinfo {year} {2001})\BibitemShut {NoStop}%
\bibitem [{\citenamefont {Coniglio}(1989)}]{Coniglio89}%
  \BibitemOpen
  \bibfield  {author} {\bibinfo {author} {\bibfnamefont {A.}~\bibnamefont
  {Coniglio}},\ }\href {\doibase 10.1103/PhysRevLett.62.3054} {\bibfield
  {journal} {\bibinfo  {journal} {Phys. Rev. Lett.}\ }\textbf {\bibinfo
  {volume} {62}},\ \bibinfo {pages} {3054} (\bibinfo {year}
  {1989})}\BibitemShut {NoStop}%
\bibitem [{\citenamefont {Fang}\ \emph {et~al.}(2020)\citenamefont {Fang},
  \citenamefont {Grimm}, \citenamefont {Zhou},\ and\ \citenamefont
  {Deng}}]{FangGrimmZhouDeng20}%
  \BibitemOpen
  \bibfield  {author} {\bibinfo {author} {\bibfnamefont {S.}~\bibnamefont
  {Fang}}, \bibinfo {author} {\bibfnamefont {J.}~\bibnamefont {Grimm}},
  \bibinfo {author} {\bibfnamefont {Z.}~\bibnamefont {Zhou}}, \ and\ \bibinfo
  {author} {\bibfnamefont {Y.}~\bibnamefont {Deng}},\ }\href@noop {} {\bibfield  {journal} {\bibinfo  {journal}
  	{{Physical Review E}}\ }\textbf {\bibinfo {volume} {102}},\ \bibinfo {pages}
  {022125} (\bibinfo {year} {2020})}\BibitemShut {NoStop}%
\bibitem [{\citenamefont {{M. Wittmann and A. P.
  Young}}(2014)}]{WittmannYoung2014}%
  \BibitemOpen
  \bibfield  {author} {\bibinfo {author} {\bibnamefont {{M. Wittmann and A. P.
  Young}}},\ }\href@noop {} {\bibfield  {journal} {\bibinfo  {journal}
  {{Physical Review E}}\ }\textbf {\bibinfo {volume} {90}},\ \bibinfo {pages}
  {062137} (\bibinfo {year} {2014})}\BibitemShut {NoStop}%
\bibitem [{\citenamefont {{J. Grimm, E. El\c{c}i, Z. Zhou, T. M. Garoni and Y.
  Deng}}(2017)}]{GrimmElciZhouGaroniDeng2017}%
  \BibitemOpen
  \bibfield  {author} {\bibinfo {author} {\bibnamefont {{J. Grimm, E. El\c{c}i,
  Z. Zhou, T. M. Garoni and Y. Deng}}},\ }\href@noop {} {\bibfield  {journal}
  {\bibinfo  {journal} {{Physical Review Letters}}\ }\textbf {\bibinfo {volume}
  {118}},\ \bibinfo {pages} {115701} (\bibinfo {year} {2017})}\BibitemShut
  {NoStop}%
\bibitem [{\citenamefont {{Z. Zhou, J. Grimm, S. Fang, Y. Deng, and T. M.
  Garoni}}(2018)}]{ZhouGrimmFangDengGaroni2018}%
  \BibitemOpen
  \bibfield  {author} {\bibinfo {author} {\bibnamefont {{Z. Zhou, J. Grimm, S.
  Fang, Y. Deng, and T. M. Garoni}}},\ }\href@noop {} {\bibfield  {journal}
  {\bibinfo  {journal} {{Physical Review Letters}}\ }\textbf {\bibinfo {volume}
  {121}},\ \bibinfo {pages} {185701} (\bibinfo {year} {2018})}\BibitemShut
  {NoStop}%
  \bibitem [{\citenamefont {Luijten}(1997)}]{luijten1997interaction}%
  \BibitemOpen
  \bibfield  {author} {\bibinfo {author} {\bibfnamefont {E.}~\bibnamefont
  		{Luijten}},\ }\emph {\bibinfo {title} {{Interaction range, universality and
  			the upper critical dimension}}},\ \href@noop {} {Ph.D. thesis, Delft University} (\bibinfo
  {year} {1997})\BibitemShut {NoStop}%
\end{thebibliography}

\end{document}